\documentclass[aps,amsmath,asssymb,superscriptaddress,groupedaddress]{revtex4}  
\usepackage{graphicx}  
\usepackage{dcolumn}   
\usepackage{bm}        
\usepackage{amssymb}   
 \usepackage{slashed}
\usepackage{epsf}
\usepackage{hyperref}
\usepackage{enumitem}
\usepackage{lipsum}
\usepackage{color}

\newcommand{\bea}{\begin{eqnarray}}
\newcommand{\eea}{\end{eqnarray}}
\newcommand{\la}{\label}
\newcommand{\be}{\begin{equation}}
\newcommand{\ee}{\end{equation}}
\newcommand{\tr}{\,\mbox{tr}\,}

\begin{document}

\title{ Spin Physics through QCD Instantons }

\author{Yachao Qian}
\address{Department of Physics and Astronomy,
Stony Brook University,  Stony Brook, NY 11794-3800.}

\author{Ismail Zahed}
\address{Department of Physics and Astronomy,
Stony Brook University,  Stony Brook, NY 11794-3800.}


\begin{abstract}
We review some aspects of spin physics where QCD  instantons play an important role.
In particular, their large contributions in semi-inclusive deep-inelastic scattering and polarized
proton on proton scattering. We also review their possible contribution in the $\mathcal{P}$-odd pion azimuthal
charge correlations in peripheral $AA$ scattering at collider energies.

\end{abstract}

\maketitle

\section{\label{sec:introduction}introduction}

Dedicated lattice simulations have revealed that the QCD  vacuum is characterized by non-trivial topological fluctuations
\cite{Leinweber:1998uu,Leinweber:2003dg}.  Instantons and anti-instantons are extrema of the 4-dimensional Euclidean action that carry
unit topological charge. They correspond to tunneling between degenerate vaccua. The instanton liquid model  with interacting
instantons and anti-instantons account for important features of the QCD vacuum, such as the spontaneous breaking of
chiral symmmetry and the large mass for the $\eta^\prime$ meson~\cite{Schafer:1996wv, nowak1996chiral}. It does not account for confinement of static charges.  Recently, it was noted that twisted instantons and anti-instantons with finite Polyakov lines preserve most of the features of the instanton liquid model and do account for confinement. 
QCD instantons may contribute substantially to small angle hadron-hadron scattering~\cite{Shuryak:2000df,Nowak:2000de,Shuryak:2003rb,Kharzeev:1999vh,Dorokhov:2004rj} 
and possibly gluon saturation at HERA~\cite{Ringwald:2000gt,Schrempp:2003tm}, as
evidenced by recent  lattice investigations~\cite{Giordano:2009vs,Giordano:2011zv}.

A number of semi-inclusive DIS experiments carried by the CLAS and HERMES collaborations 
~\cite{Airapetian:2004tw,Airapetian:2009ae,Avakian:2010ae}, and more recently with polarized
protons on protons by the STAR and PHENIX collaborations~\cite{Abelev:2008af,Eyser:2006bc, Adams:1991cs},
have revealed large spin asymmetries  in polarized lepton-hadron and hadron-hadron
collisions at collider energies.  These effects  are triggered by $\mathcal{T}$-odd contributions in the
scattering amplitude.   Perturbative QCD does not support the $\mathcal{T}$-odd contributions,  which are usually
parametrized in the initial state (Sivers effect)~\cite{Sivers:1989cc,Sivers:1990fh} or 
the final state (Collins effect)~\cite{Collins:1992kk,Collins:1993kq}. Non-perturbative
QCD with instantons allow for large spin asymmetries as discussed by Kochelev
and others~\cite{Kochelev:1999nd,Dorokhov:2009mg,Ostrovsky:2004pd,Qian:2011ya}.  In~\cite{Kochelev:1999nd} a particularly large Pauli form factor was noted, with  an important contribution to the Single Spin Asymmetry (SSA) in polarized proton on proton scattering.

In this paper, we review some recent developments regarding our understanding of spin physics in the instanton liquid
model. Assuming that the vacuum is populated by semi-classical but interacting instatons and anti-instantons, with
the vacuum parameters fixed by the spontaneous breaking of chiral symmetry in bulk, we explicit their effects on
semi-inclusive DIS processes as well as singly polarized $pp$ scattering. In both cases, uncorrelated instantons or anti-instatons
are at work. We  show that the effects of correlations between instantons and anti-instantons through fluctuations are also
important in both doubly polarized $pp$ scattering as well as through $\mathcal{P}$-odd effects in peripheral $AA$ scattering. 

The paper is organized as follows: in section II we detail the role of a single instanton and anti-instanton on the 
single spin assymetry in semi-inclusive DIS scattering and in polarized $pp$ scattering. The large instanton contributions
appear to be supported by current experiments. In section III we discuss the role of local fluctuations in the compressibility
as well as the topological susceptibility  in doubly polarized $pp$ and peripheral $AA$ scattering. Our conclusions and
prospects follow in section IV. In Appendix A we detail the large instanton vertex contribution to both the electromagnetic
and chromo-magnetic interactions with the corresponding large magnetic moments.

\section{\label{sec:spinandinstanton}Spin effects through one instanton}

To best illustrate the important role played by instantons in QCD spin physics, consider a light quark in the fundamental 
color representation propagating in an external SU(2) colored Yang-Mills gauge field with a chromo-magnetic  field ${\bf B}$ 
and a chromo-electric  field ${\bf E}$ field. Generically~\cite{Shuryak:2004tx}

\be
\left (-\nabla^2+4g_s\,{\bf S}\cdot ({\bf B}\mp {\bf E})\right)\varphi^\pm=0
\label{DIR}
\ee
with $i\nabla=i\partial  +A$ and     ${\bf S}^a$ the SU(2) spin generators. The signs in (\ref{DIR}) refer to 
the chirality of the quark. Large quark amplitudes as  polarized zero modes occur when the spin contribution
 (second term) balances the squared kinetic contribution (first term) in (\ref{DIR}). For a self-dual instanton
 with ${\bf B}={\bf E}$ the negative chirality quark produces a large zero mode state through the magnetic 
 moment term

\be
\left (-\nabla^2+4g_s\,{\sigma}\cdot {\bf B}\right)\varphi_D^-=0
\label{DIR1}
\ee
and similarly for an anti-self-dual anti-instanton. Typically ${\bf E}, {\bf B}\approx 1/g_s\rho^2$ with $\rho\approx 1/3$ 
fm the instanton  or anti-instanton size in the vacuum and $g_s$ the strong gauge coupling. So the induced and large magnetic moment 
in (\ref{DIR1}) is about $\mu_D\approx {\bf n}\rho^4$ where ${\bf n}\approx 1\,{\rm fm}^{-4}$ is the density of instantons in the vacuum
~\cite{Schafer:1996wv,nowak1996chiral}. In contrast, perturbative QCD generates small magnetic moments or $\mu_{PT}\approx g_s$. 

In a similar way, a propagating gluon in an external SU(2) colored SU(2) gauge field acquires also an effective
and large magnetic moment. Indeed, the analogue of (\ref{DIR}) for 
a massless gluon in a covariant (Feynman0) background gauge  $\nabla^\mu {\bf a}_\mu=0$ is

\be
\left(-\nabla^2\delta_{\mu\nu} -2ig_sF_{\mu\nu}\right){\bf a}^\nu=0
\label{G1}
\ee
The colored gluon in (\ref{G1}) has two physical polarizations as both the longitudinal
and time-like are gauge artifacts. Using the decomposition ${\bf a}_\mu={\bf e}^a_\mu{\bf \Psi}^a$
with $a$ transverse we have

\be
\left(-\nabla^2+\frac {g_s}2 \Sigma^{\mu\nu}F_{\mu\nu}\right){\bf \Psi}^a=0
\label{G2}
\ee
with $i\Sigma_{\mu \nu}=4{\bf e}_\mu^T{\bf e}$ playing the role of the spin in the gluon transverse polarization space. 
(\ref{G2}) is the analogue of (\ref{G1}) with an induced and large magnetic moment $\mu_G\approx {\bf n}\rho^4$ as
well.

In the Appendix we give a quantitative derivation of these estimates. 
These semi-classical and large spin effects will now be explored in
processes with polarized protons and in peripheral $AA$ collisions
sensitive to $\mathcal{P}$-odd fluctuations, in the framework  of the instanton liquid
model.

\subsection{\label{subsec:SSASIDIS}Single Spin Asymmetry Semi-Inclusive Deep Inealstic Scattering }

\begin{figure}
\includegraphics[height=50mm]{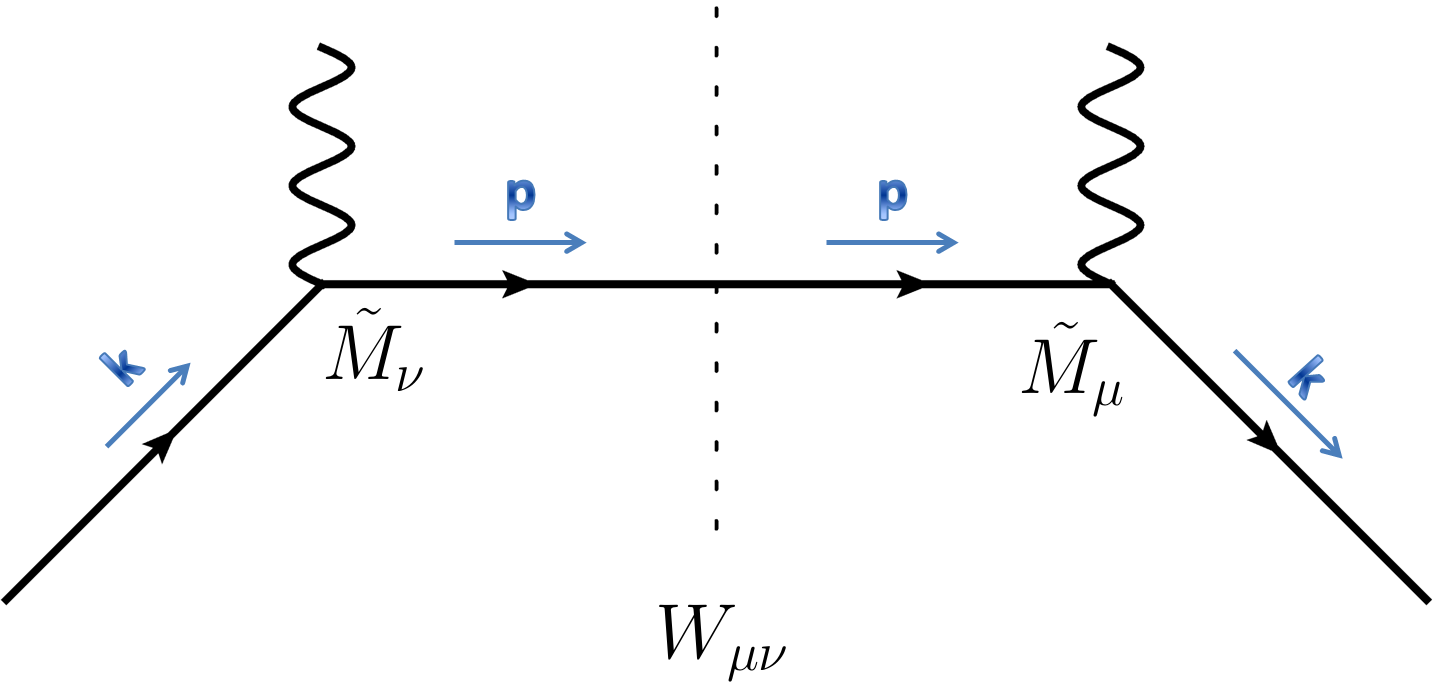}
\caption{ $l$ and $l^\prime$ denote the momentum for the incoming and outgoing lepton. $p$ and $k$ are the momenta of the incoming and outgoing quark. The lepton and the quark exchange one photon in the single instanton background. }
\label{CARTOONDIS}
\end{figure}

 
To set up the notations for the semi-inclusive processes in deep inelastic scattering, we consider a proton at
rest in the LAB frame with transverse polarization as depicted in Fig.~\ref{CARTOONDIS}. 
The incoming and outgoing leptons are unpolarized. The polarization of the target proton in relation to the DIS
kinematics is shown in Fig.~\ref{dip3dgraph}.  Throughout, the spin dependent asymmetries will be evaluated 
at the partonic level. Their conversion to the hadronic level will follow the qualitative arguments presented in~\cite{Kochelev:1999nd,Dorokhov:2009mg,Ostrovsky:2004pd}.

In general, the spin averaged leptonic tensor reads

\bea\la{leptontensor}
L^{\mu \nu} = \frac{1}{2} \tr[\slashed{l}^\prime \gamma_\mu \slashed{l} \gamma_\nu]
\label{LEPTONIC}
\eea
while the color averaged hadronic tensor in the one instanton background reads

\bea\label{hadrontensor}
W_{\mu \nu} = \sum_{color} \frac{1}{2} \tr[\slashed{k} \tilde{M}_\mu \slashed{p}(1 + \gamma_5 \slashed{s}) \gamma_0 \left(\tilde{M}_\nu\right)^\dagger \gamma_0] 
\label{HADRONIC}
\eea
with the constituent vertex 

\be
\tilde{M}_\mu = \gamma_\mu + \tilde{M}_\mu^{(1)}
\ee
that includes both the perturbative $\gamma_\mu$  and the non-perturbative insertion $M_\mu^{(1)}$. 
In the Appendix we detail its derivation following the 
original arguments in~\cite{Moch:1996bs, Ostrovsky:2004pd, Qian:2011ya}. After color averaging, the result is

\bea\la{dipm1}
{\rm Im} \left(  \left< \tilde{M}_\mu^{(1)}  \right> \right) = -  \frac{4 \pi^2 \rho^2}{\lambda_* Q^2}[ \gamma_\mu \slashed{k} + \slashed{p} \gamma_\mu] (1 - f(\rho Q)) 
\label{1VERTEX}
\eea
with $f(a) = a K_1(a)$ with $K_1$ a modified Bessel function. Here $\rho$ is the instanton size and
$\lambda_* \approx 1/(0.2  {\rm  GeV})^3 $~\cite{Schafer:1996wv,nowak1996chiral} 
is a typical near-mode in the zero-mode-zone as discussed in the Appendix. 
The normalized lepton-hadron cross section of Fig.\ref{CARTOONDIS} follows in the form

\bea\la{crosssection}
\frac{d \sigma}{dx dy dz d\phi} = y \frac{\alpha^2}{  Q^6} L^{\mu \nu} W_{\mu \nu} \sum_i e_i^2 f_i (x, Q^2) D_i (z)
\label{CROSS}
\eea
with $y = P \cdot q / P \cdot l $,  where $e_i$ is the i-parton electric charge, $f_i$ its momentum fraction distribution and $D_i$ its framentation function.

The perturbative contribution to the hadronic tensor follows from $M_\mu\rightarrow \gamma_\mu$,

\bea 
W_{\mu \nu}^{(0)} =  \frac{N_c}{2} \tr[\slashed{k} \gamma_\mu \slashed{p} \gamma_\nu] 
\label{HADRONIC}
\eea
Thus the leading perturbative contribution

\be\la{zeroorder}
\frac{d^{(0)} \sigma}{dx dy dz d\phi} =2 N_c  \frac{\alpha^2}{Q^2} \frac{1+ (1-y)^2}{y} \sum_i e_i^2 f_i (x, Q^2) D_i (z)
\ee
with $N_c$ the number of colors. The sum is over the charges $e_i$ of the quarks.
The non-perturbative instanton contribution to (\ref{CROSS})
is a cross contribution in the hadronic tensor in (\ref{CROSS}) after inserting the one-instanton vertex (\ref{1VERTEX})

\bea
\label{WXX}
\left<W_{\mu \nu}^{(1)} \right> &=&    i \frac{2 \pi^2 \rho^2}{\lambda_* Q^2} [ 1 - f(\rho Q) ] \left(   \tr[\slashed{k}  \gamma_\mu  \slashed{p}  \gamma_5 \slashed{s}  ( \gamma_\nu \slashed{p} + \slashed{k} \gamma_\nu )]     - \tr[\slashed{k} (\gamma_\mu \slashed{k} + \slashed{p} \gamma_\mu) \slashed{p}  \gamma_5 \slashed{s}\gamma_\nu]    \right)\\
 &=&  -   \frac{16 \pi^2 \rho^2}{\lambda_* Q^2} (1 - f(\rho Q)) (p+k)_{\{ \mu} \epsilon_{\nu \} a b c}s^a k^b  p^c   \nonumber
\eea
where $p \cdot s = 0$ and  the short notation $(\cdots)_{\{ \mu} \epsilon_{\nu \} a b c} \equiv (\cdots)_\mu \epsilon_{\nu a b c}  + (\cdots)_\nu \epsilon_{\mu a b c}$
is used.    If we set $p=xP$ and $k={K}/{z}$ and note that $p+k = 2 p + q$, then (\ref{WXX}) simplifies
\be\la{hadrontensor2}
\left< W_{\mu \nu}^{(1)} \right>=  - \frac{2^5 \pi^2 \rho^2}{\lambda_* Q^2} \frac{x^2}{z} (1 - f(\rho Q)) (P + \frac{q}{2x})_{ \{ \mu} \epsilon_{ \nu \} a b c}s^a K^b  P^c 
\ee
Combining (\ref{leptontensor}) and (\ref{hadrontensor2}) yields
\be \la{lwdip}
\left< W_{\mu \nu}^{(1)}L^{\mu \nu} \right> = - \frac{2^7 \pi^2 \rho^2}{ \lambda_* Q^2} \frac{x^2}{z} (1 - f(\rho Q))M \left(  E  \epsilon_{\nu a b c} l^{\prime \nu} s^a K^b  P^c +   E^\prime   \epsilon_{\nu a b c} l^{ \nu} s^a K^b  P^c \right)
\ee
where $E$ ($E'$) is the energy of the incoming (outgoing)  (anti)electron.
The leading instanton contribution to the total cross section (\ref{CROSS}) follows by inserting (\ref{lwdip}) into (\ref{HADRONIC}).

\be \label{SIGMA1}
\frac{d^{(1)} \sigma}{dx dy dz d\phi} =  \Delta_\perp q_q  \frac{\alpha^2}{y Q^2}  \sum_q \frac{64 \pi^2 \rho^2   e_q^2  D_q(z)}{ \lambda_* Q } \frac{K_\bot}{z Q}  (1 - f(\rho Q))   \frac{1-y-x^2 y^2 \frac{M^2}{Q^2}}{\sqrt{1 + 4 \frac{M^2}{Q^2}x^2}}  \sin(\phi - \phi_s)   
\ee 
where  $\Delta_\perp q_q (x, Q^2) = s_\perp f_{q}(x) $ is the spin polarized distribution function for the quark in the transversely polarized proton.  The overall sign in (\ref{SIGMA1}) is tied with the conventional sign of the proton mass $M$.

\begin{figure}[b]
\includegraphics[height=43mm]{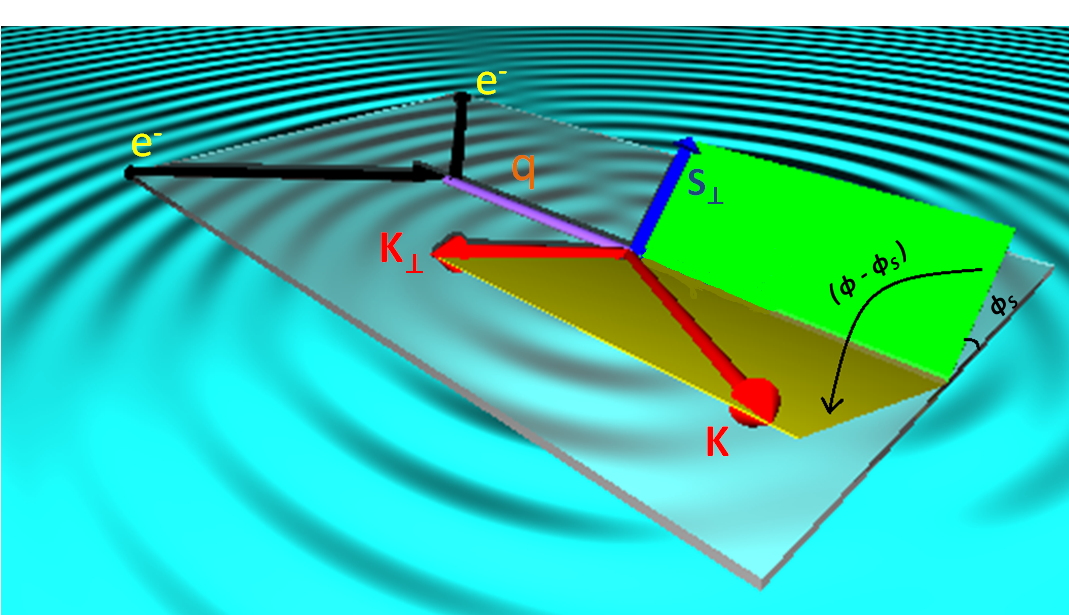}
\caption{\la{dip3dgraph}   The lepton and photon are in the same plane. The angle between the transversely polarized spin $s_\perp$ and this plane is $\phi_s$. The angle between the transversely spatial  momentum $K_\perp$ of the outgoing pion and the plane is $\phi$. }
\end{figure}

To compare  with experiment, we will use the spin structure function~\cite{Airapetian:1998wi}
\be
  g_1(x,Q^2) = \frac{1}{2} \sum_q e_q^2 (\Delta q_q (x,Q^2) + \Delta \bar{q}_q (x,Q^2))  
\ee
Since we are only interested in the SSA in hard scattering processes, we set $D_q (z) = 1$.  A comparison of (\ref{zeroorder}) with (\ref{SIGMA1}), yields for the SSA

\be \la{dipsindiff}
A^{{\rm sin}(\phi - \phi_s)}_{UT} = \frac{32 \pi^2 \rho^2}{N_c Q \lambda_*}  (1 - f(\rho Q)) \frac{ K_\bot}{zQ}  
   \frac{1 }{ 1 + (1-y)^2} \frac{1-y-x^2 y^2 \frac{M^2}{Q^2}}{\sqrt{1 + 4 \frac{M^2}{Q^2}x^2}}      \frac{g_1}{F_1} 
\ee 
with $F_1(x) = \frac{1}{2} \sum_{q} e_q^2 f_q(x)$ .  This is usually referred to as the Sivers contribution.   
In Fig.~\ref{autnew} we compare \eqref{dipsindiff} to the results reported by HERMES~\cite{Airapetian:2009ae}.
We use a direct probing of the dependence of the transverse
spin asymmetry on $x,z,K_\perp$.  For instance, take the $x$ dependent asymmetry of $\pi^-$ with
the empirical parametrizations to fit the reported kinematics from HERMES~\cite{Airapetian:2009ae}:
 $<y> = -95.737 x^3 + 52.459 x^2 - 9.0816 x + 0.9495$, $<z>= 15.67 x^3 - 8.8459 x^2 + 1.5193 x + 0.2884$, $<K_\bot> = 665.15 x^4 - 444.02 x^3 + 105.99 x^2 - 10.843 x + 0.7502 \,({\rm GeV})$ , and $<Q^2> = 20.371 x + 0.4998 ({\rm GeV^2})$. $R^2$ is above 0.97 for all the parametrizations.

\begin{figure*}
\includegraphics[width = 120mm]{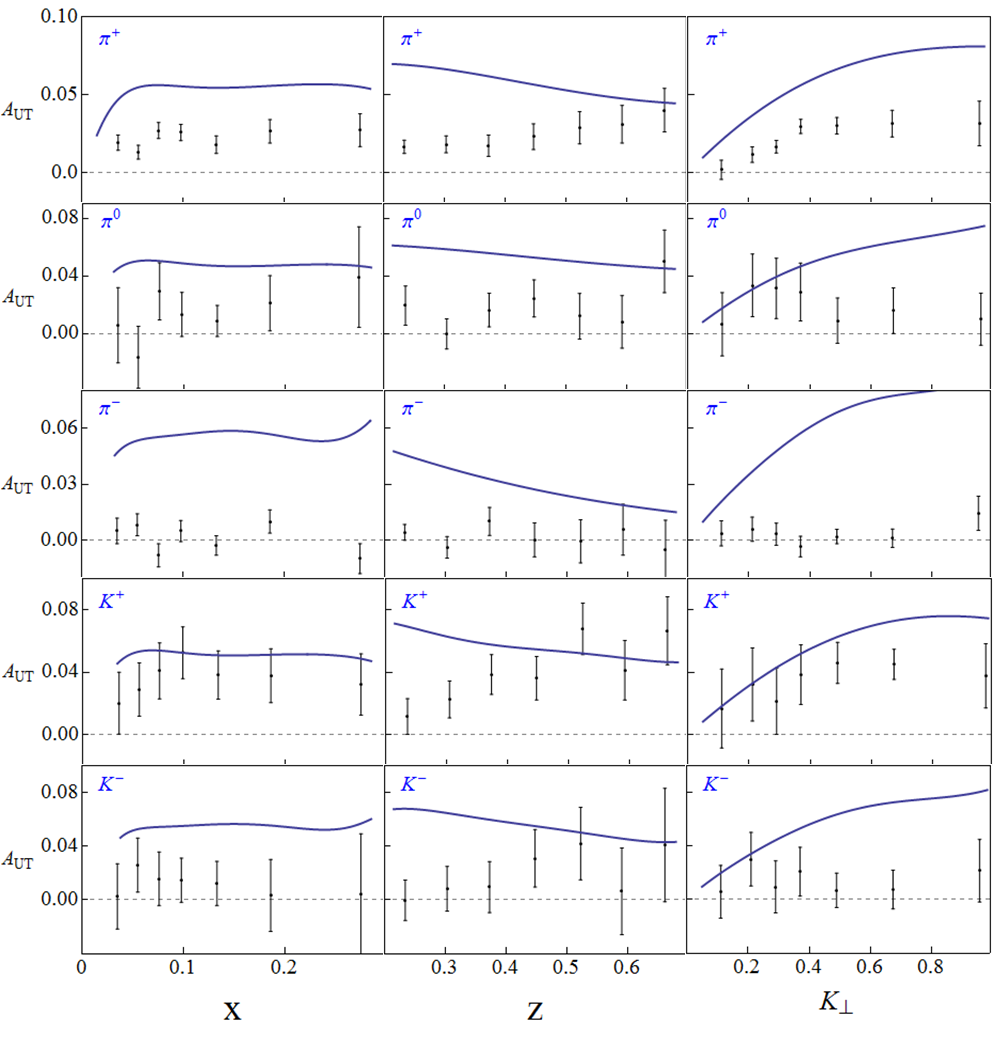}
\caption{\la{autnew} Transversly polarized spin asymmetry (solid line) versus data~\cite{Airapetian:2009ae}.}
\end{figure*}

\subsection{\label{subsec:singlespinresultpp} Sinlge Spin Asymmetry in $pp$--- Transverse parton distribution function}



\begin{figure}[!htb]
\minipage{0.45\textwidth}
\includegraphics[height=35mm]{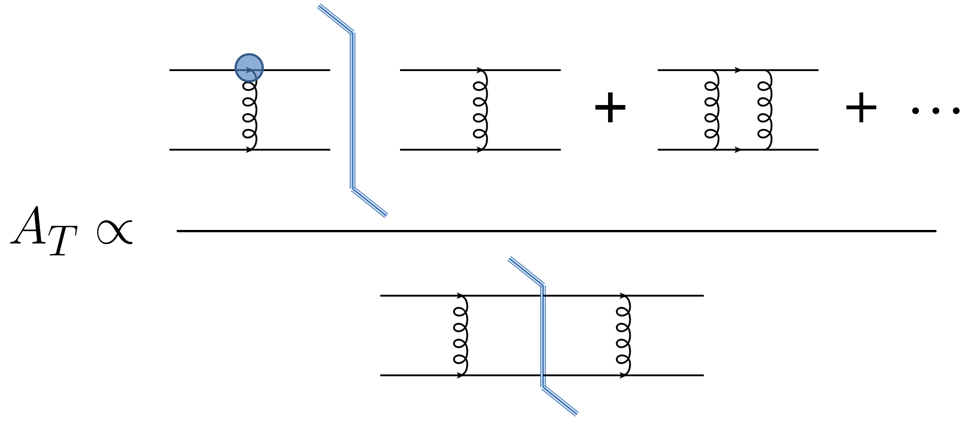}
\caption{Schematically diagrammatic contributions to the SSA through the Pauli Form factor~\cite{Kochelev:2013zoa} \la{ssaexpansion} }
\endminipage\hfill
\minipage{0.45\textwidth}
\includegraphics[height=35mm]{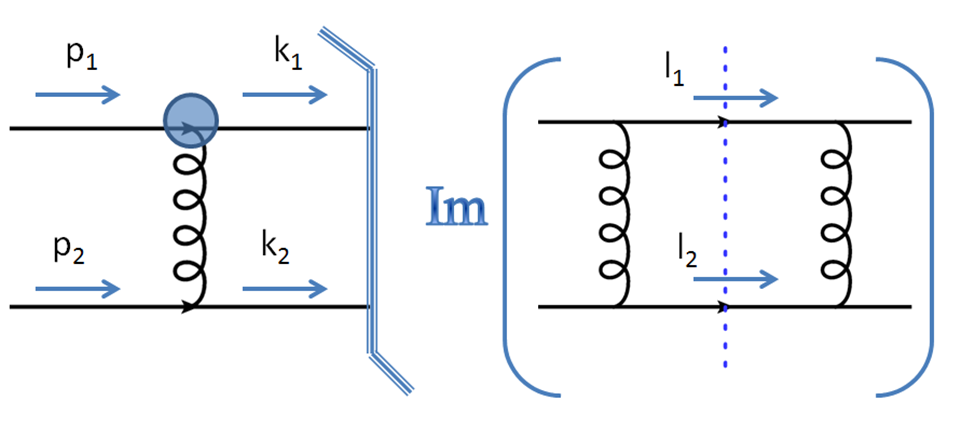}
\caption{ Leading diagrammatic contribution to the SSA through the Pauli form factor.  \la{ssaimaginary} }
\endminipage
\end{figure}

In this section we briefly review 
the SSA in semi-inclusive and polarized $p_\uparrow p\rightarrow \pi^{\pm, 0} X$ experiments, following the recent analysis 
in~\cite{Kochelev:2013zoa,Qian:2014aza}.
In going through an instanton, the chirality of the light quark can be flipped as we noted in (\ref{DIR1}). Using the Pauli form factor 
discussed in the Appendix, the SSA follows from the diagrams of Fig.~\ref{ssaexpansion}. 
As noted in \cite{Kochelev:2013zoa}, the leading diagram contributing to the SSA is displayed in Fig.~\ref{ssaimaginary}. 
Note that Fig.~\ref{ssaimaginary} is of the same order in $g_s$ as the zeroth order diagram in Fig.~\ref{ssaexpansion},  since
the chirality-flip effective vertex (Eq.~\ref{effectivevertex}) is semi-classical and of order $1/g_s^2$. 
The zeroth order differential cross section reads

\be\la{zeroorder}
d^{(0)} \sigma  \sim     \frac{64 g_s^4}{|p_1 - k_1|^4} [(k_1 \cdot p_2) (k_2 \cdot p_1) + (k_1 \cdot k_2) (p_1 \cdot p_2)]
\ee
The  first order differential cross section for the chirality flip reads~\cite{Potter:1997za}
\be\la{firstordereq1}
d^{(1)} \sigma \sim  i \frac{g_s^6}{(k_1 - p_1)^2 } \frac{1   }{ 16\pi}  \frac{(4 \pi)^\epsilon}{\Gamma(1 - \epsilon)} \frac{\mu^{2 \epsilon}}{s^\epsilon} \int_0^1 d y~[y (1-y)]^{- \epsilon} \int_0^{2 \pi}\frac{ d \phi_l}{2 \pi}  \frac{1}{(l_1 - k_1)^2 } \frac{1}{(p_1 - l_1)^2 }\mathcal{G} (\Omega)
\ee
where  $y = (1+\cos \theta_l)/2$, $\pm \theta_l$ is the longitudinal angle of $l_{1/2}$ and
\be
\mathcal{G} (\Omega) \equiv  \tr[ (M_\mu^a)^{(1)}  \slashed{p}_1  \gamma_5 \slashed{s} \gamma_\nu t^b \slashed{l}_1 \gamma_\rho  t^c \slashed{k}_1  ] \tr [   \gamma^\mu t_a \slashed{p}_2 \gamma^\nu t_b \slashed{l}_2  \gamma^\rho t_c \slashed{k}_2] 
\ee
From Sec-\ref{subsec:type2} in the Appendix we have 
\be 
\left<(M_\mu^a)^{(1)} \right> =  - t^a     \sigma_{\mu\nu} q^\nu    \Psi  
\ee
where  
\be
\Psi =   \frac{   F_g(\rho_c\, Q)   \pi^4   (n_I \rho^4_c)   }{m_q^* g_s^2}
\ee

To simplify the analysis and compare to the existing semi-inclusive data, we use the  kinematics 
\bea\la{kinematicssimple}
 p_{1/2} &=&  \frac{\sqrt{\tilde{s}}}{2} (1, 0, 0, \pm 1) \nonumber\\
k_{1/2} &=&   \frac{\sqrt{\tilde{s}}}{2}  (1,  \pm \sin \theta  \sin \phi  ,  \pm \sin \theta   \cos \phi  , \pm \cos \theta  ) \nonumber\\
s  &=& (0,0, s^\perp , 0) 
\eea
where   $\sqrt{\tilde{s}}$ is the total energy of the colliding "partons". It is  simple to show that $d^{(1)} \sigma \sim \vec{k}_1 \cdot ( \vec{p}_1 \times \vec{s}) \sim \sqrt{\tilde{s}} s^\perp k^\perp_1 \sin \phi$, which results in SSA. For simplicity, we calculate the first differential cross section $d^{(1)} \sigma$ with $\phi = \pi/2$, where the transverse momentum of the outgoing particle lines along the $x$ axis.  Straightforward algebra yields
\bea
\left< d^{(1)} \right> \sigma &\sim&  s^\perp k^\perp_1 \frac{2 g_s^4 }{3 \pi} \frac{\Gamma(- \epsilon)}{\Gamma(2 - 2\epsilon) \Gamma(1 - \epsilon)}   \csc^2 (\theta)   (4 \pi)^\epsilon  \frac{\mu^{2 \epsilon}}{s^\epsilon}
\left( \Psi g_s^2 \right)   \nonumber\\
&& \times [25 \epsilon - 12 + \cos \theta (\epsilon (9 + 2  \epsilon) - 4)   {}_2F_1 (1, 1- \epsilon, 1 - 2 \epsilon, \sec^2 \frac{\theta}{2} )   + \epsilon (1 - \cos \theta) {}_2F_1 (2, 1- \epsilon, 1 - 2 \epsilon, \sec^2 \frac{\theta}{2} )    ]\nonumber\\
\eea
where ${}_2F(a,b,c; y)$ is a  hypergeometric function. We note that $ |{}_2F_1 (1,1,1;y) |$ is much larger than $| {}_2F_1^{(0,1,0,0)}(1,1,1;y)|$ and $|{}_2F_1^{(0,0,1,0)}(1,1,1;y)|$  for $y \sim1$. Therefore
\be \la{typetworesult}
\left< d^{(1)} \sigma  \right>    \sim     s^\perp k^\perp_1 \frac{2 g_s^4 }{3 \pi} \left( \Psi g_s^2 \right)   \csc^4 ( \frac{\theta}{2})  (3 + \cos \theta ) \left( - \frac{1}{\epsilon} + 2 \gamma_{\rm E} + \ln (\frac{\tilde{s}}{4 \pi \mu^2})  \right) 
\ee 
The divergence in (\ref{typetworesult})  stems from the exchange of soft gluons in the box diagram.
In~\cite{Kochelev:2013zoa} it was regulated using a constituent gluon mass $m_g$. For $\theta_l \sim 0$,   $\vec{l}_1$ is parallel to   $\vec{p}_1$, and this collinear divergence could be regulated by restricting $- (l_1 - p_1)^2 > m^2_g$ or equivalently setting $y_{\rm max} \sim 1-  c\, {m_g^2}/{\tilde{s}}$ with
$c$ an arbitrary constant of order 1. This regularization amounts to the substitution

\be\la{reducedinte}
\int_0^1 dy \longrightarrow \left( \int_0^{\frac{1 + \cos \theta}{2} - c \frac{m_g^2}{\tilde{s}}}  +\int_{\frac{1 + \cos \theta}{2}+ c \frac{m_g^2}{\tilde{s}}}^{1- c \frac{m_g^2}{\tilde{s}}} \right)   dy
\ee
in Eq.~\ref{firstordereq1},
where we have also regulated the collinear divergence when  $\vec{l}_1$ is parallel to $\vec{k}_1$. Thus

\be
 \left( - \frac{1}{\epsilon} + 2 \gamma_{\rm E} + \ln (\frac{\tilde{s}}{4 \pi \mu^2})  \right)  \longrightarrow  \ln \left( c \frac{\tilde{s}}{m_g^2} \right)    + \ln\left( \frac{1 - \cos \theta}{1 + \cos \theta} \right)
\ee
The regulated SSA is now given by

\be
A_{T}^{\sin \phi} \approx \left< \frac{  d^{(1)} \sigma  }{    d^{(0)} \sigma  } \right> = s^\perp k^\perp_1 \left( \frac{\Psi g_s^2}{\pi} \right)  \frac{      (3 + \cos \theta )     }{6  (5 + 2 \cos \theta + \cos^2 \theta)} \left[  \ln \left(   c \frac{\tilde{s}}{m_g^2} \right)    + \ln\left( \frac{1 - \cos \theta}{1 + \cos \theta} \right)    \right]
\ee
where the zeroth order cross section in Eq.~\ref{zeroorder} is used for normalization.


To compare with the semi-inclusive data on $p_\uparrow p\rightarrow \pi X$,  we set $s^\perp u(x,Q^2) = \Delta_s u (x,Q^2)$ and $s^\perp d(x,Q^2) = \Delta_s d (x,Q^2)$, with $\Delta_s u (x,Q^2)$ and $\Delta_s d (x,Q^2)$ as the spin polarized distribution functions of the valence up-quarks and valence down-quarks in the proton respectively.    For forward $\pi^+$, $\pi^-$ and $\pi^0$ productions, the SSAs are

\be\la{piplus}
 A_{T}^{\sin \phi} (\pi^+) =    k^\perp  \frac{\Delta_s u(x_1  ,Q^2)}{u(x_1  , Q^2)}  \left( \frac{\Psi g_s^2}{\pi} \right)\frac{   (3 +\cos \theta)   }{6  (5 + 2\cos \theta  + \cos^2 \theta)}\left[  \ln \left(   c \frac{\tilde{s}}{m_g^2} \right)    + \ln\left( \frac{1 - \cos \theta}{1 + \cos \theta} \right)    \right]
\ee

\be\la{piminus}
 A_{T}^{\sin \phi} (\pi^-) =   k^\perp  \frac{\Delta_s d(x_1 ,Q^2)}{d(x_1  , Q^2)}  \left( \frac{\Psi g_s^2}{\pi} \right)\frac{   (3 +\cos \theta)   }{6  (5 + 2\cos \theta  + \cos^2 \theta)}\left[  \ln \left(   c \frac{\tilde{s}}{m_g^2} \right)    + \ln\left( \frac{1 - \cos \theta}{1 + \cos \theta} \right)    \right]
\ee

\be\la{pizero}
 A_{T}^{\sin \phi} (\pi^0) =    k^\perp   \frac{\Delta_s u(x_1  ,Q^2) + \Delta_s d(x_1 ,Q^2)}{u(x_1  , Q^2) + d(x_1  , Q^2)}   \left( \frac{\Psi g_s^2}{\pi} \right) \frac{   (3 +\cos \theta)   }{6  (5 + 2\cos \theta  + \cos^2 \theta)}\left[  \ln \left(   c \frac{\tilde{s}}{m_g^2} \right)    + \ln\left( \frac{1 - \cos \theta}{1 + \cos \theta} \right)    \right]
\ee 
According to \cite{Hirai:2006sr, Adams:1997dp}
\bea
  \frac{\Delta_s u(x  ,Q^2)}{u(x  , Q^2)} &=& 0.959 - 0.588 (1 - x^{1.048}) \nonumber\\
   \frac{\Delta_s d(x  ,Q^2)}{d(x  , Q^2)} &=& - 0.773 + 0.478 (1- x^{1.243})\nonumber\\
 \frac{u(x  , Q^2)}{d(x  , Q^2)} &=& 0.624 (1-x)
\eea

\begin{figure}[!htb]
\minipage{0.48\textwidth}
\includegraphics[height=40mm]{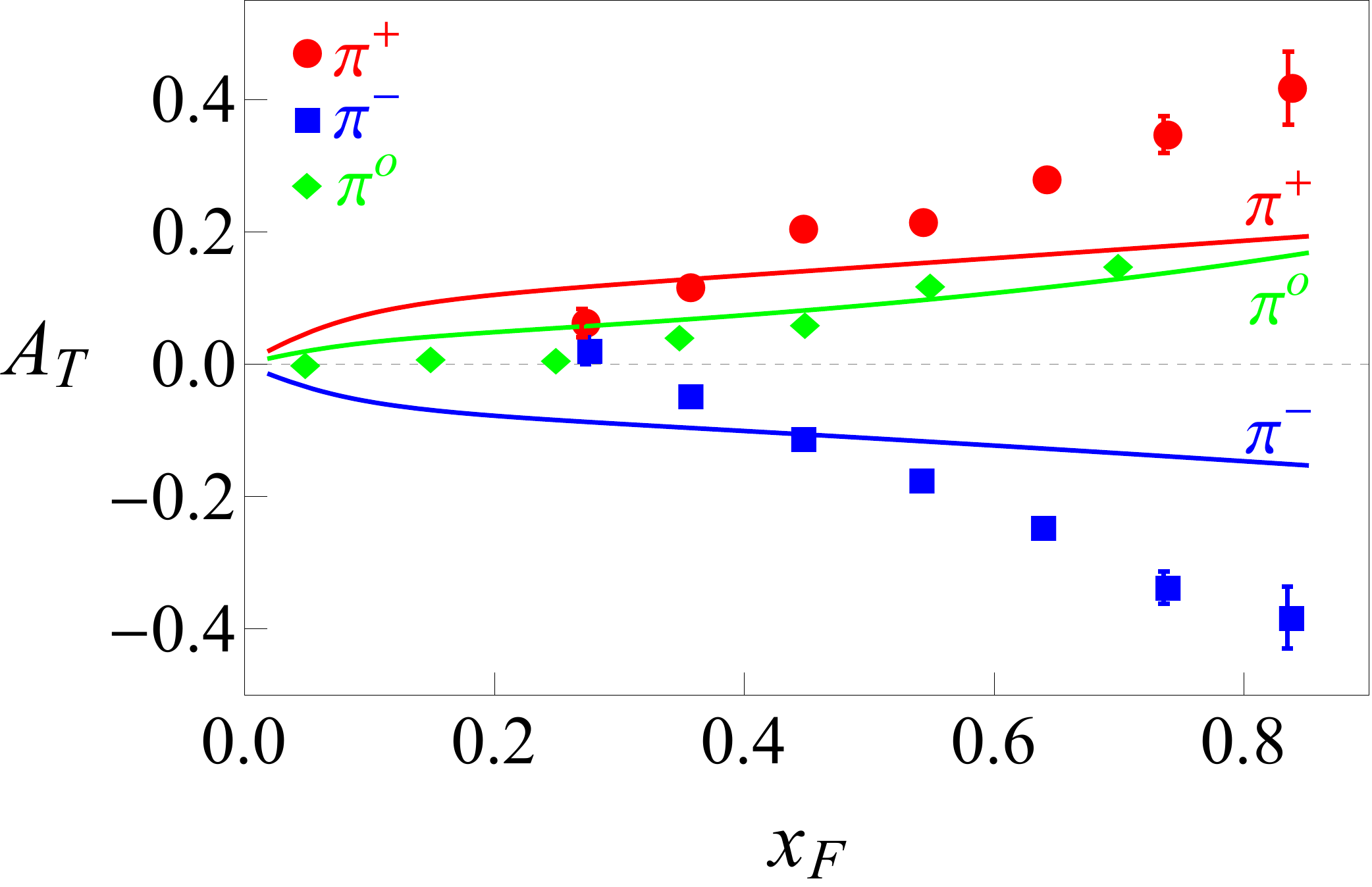}
\endminipage\hfill
\minipage{0.48\textwidth}
\includegraphics[height=40mm]{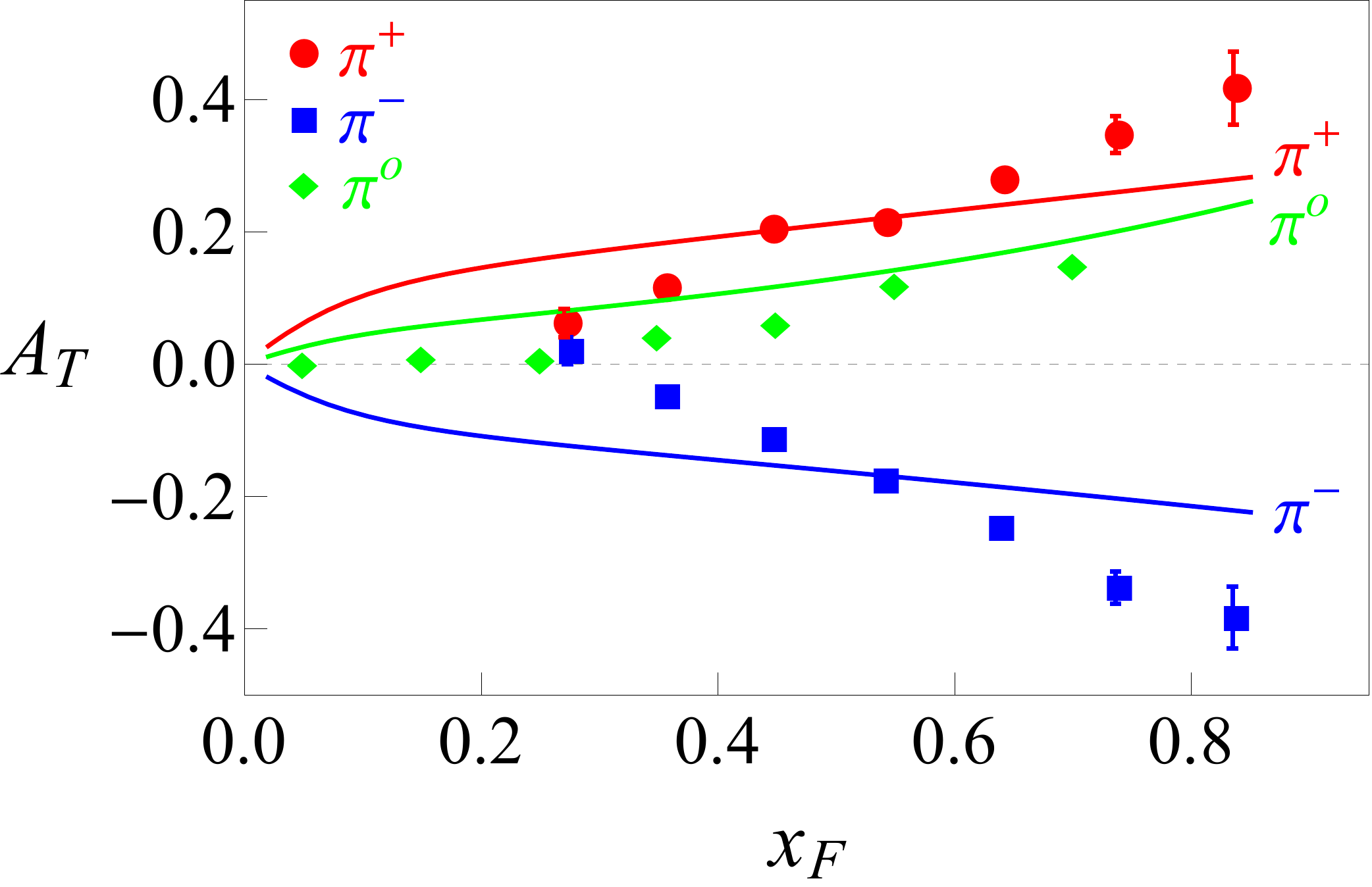}
\endminipage
  \caption{ $x_F$ dependent SSA in $p_\uparrow p\rightarrow \pi X$ collisions at $\sqrt{s}=19.4 {\rm GeV}$~\cite{Skeens:1991my}.  The solid lines are the analytical results in Eq.~\ref{piplus}- Eq.~\ref{pizero} with $c = 2$ (left). A regulator using a massive gluon propagator
yields the results  (right). }\label{ssafigure}
\end{figure}

\begin{figure}[!htb]
\minipage{0.48\textwidth}
\includegraphics[height=40mm]{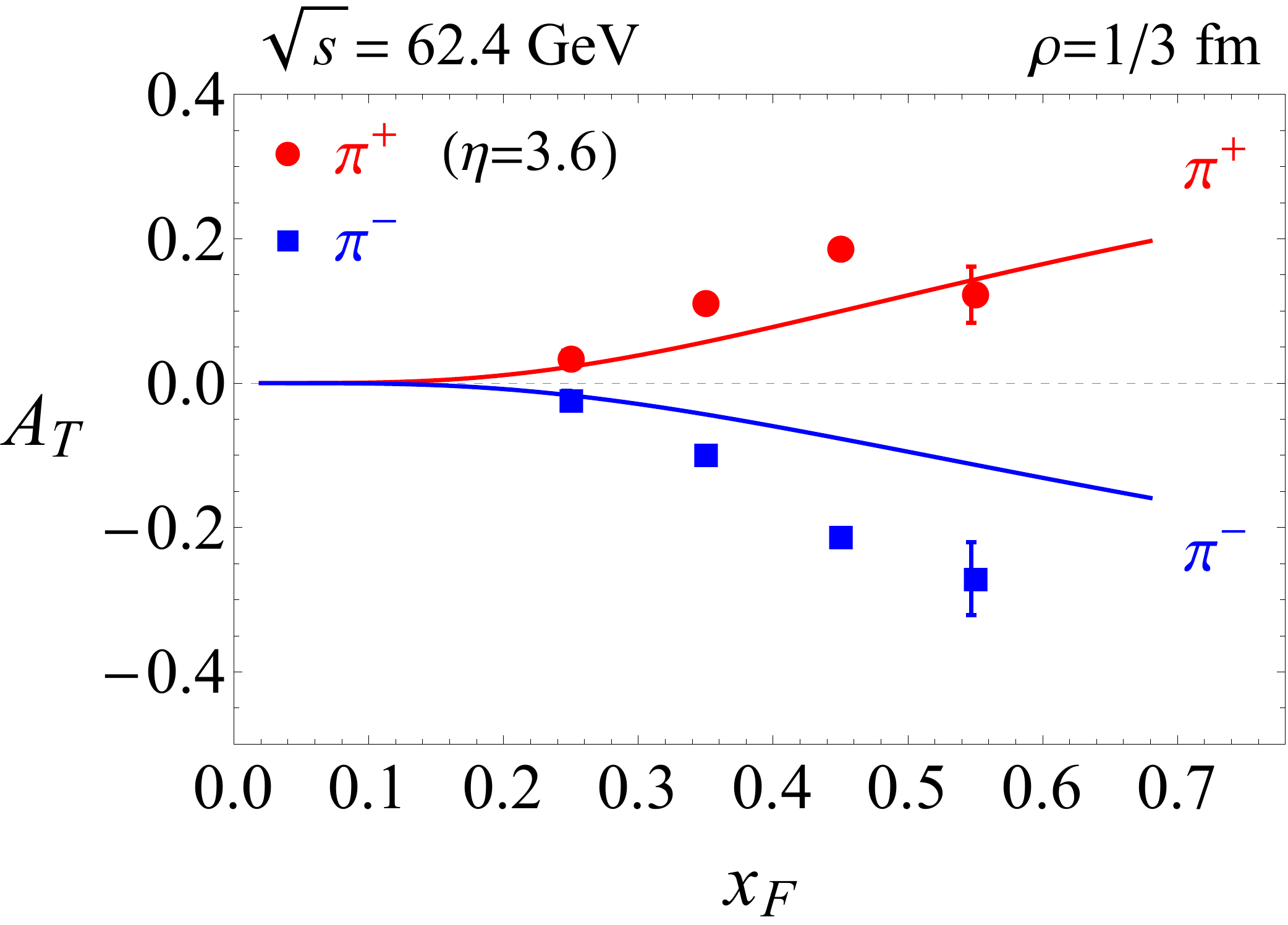}
\endminipage\hfill
\minipage{0.48\textwidth}
\includegraphics[height=40mm]{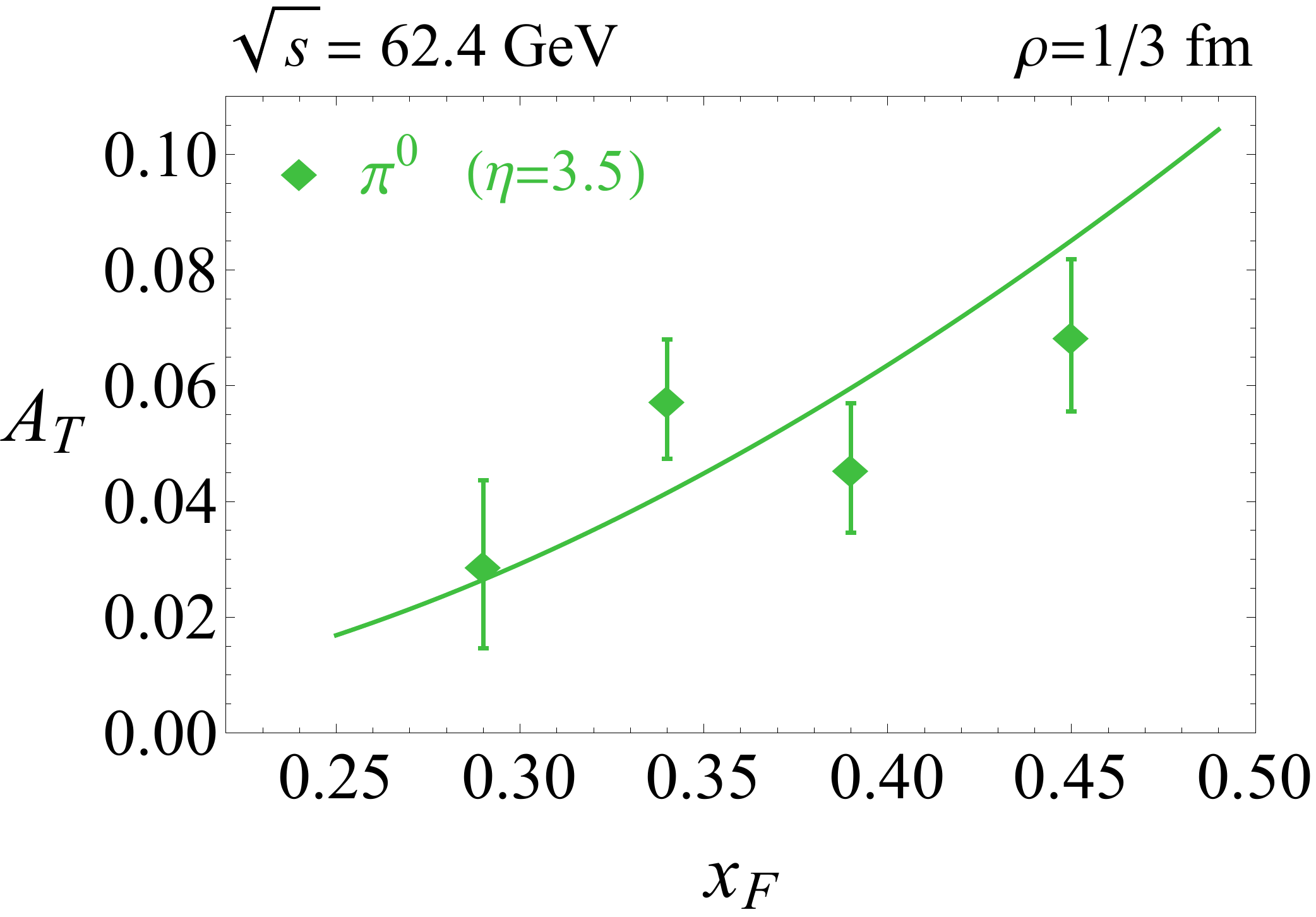}
\endminipage
  \caption{ $x_F$ dependent SSA in $p_\uparrow p\rightarrow \pi X$ collisions at $\sqrt{s}=62.4\,{\rm GeV}$.   Data (left) is from~\cite{Arsene:2008aa}. Data (right) is from~\cite{Adare:2013ekj}.}\label{ssafigure2}
\end{figure}

These results can be compared to the experimental measurements in~\cite{Skeens:1991my}. For simplificty,  we assume the same fraction for each proton $\left<x_1\right> = \left<x_2\right>  = \left<x \right> $,  and $\left<k^\perp \right> \approx \left<K_\perp\right>$ is the transverse momentum of the outgoing pion. We then have $ \sqrt{s} \left<x\right> \left<\sin \theta \right>   = 2 \left<K_\perp\right>$ and $\left<x \right> \left<\cos\theta \right> = \left<x_F\right>$.   For large $\sqrt{s}$, we also have $\left<Q\right> \approx \left<K_\perp \right> \sqrt{\left<x \right>/ \left<x_F \right>}$. We set $c = 2$ and $\left< K_\perp \right> = 2 {\rm GeV}$ for the outgoing pions.  $n_I \approx 1/{\rm fm}^4$ is the effective instanton density, $\rho \approx 1/3 {\rm fm}$ the typical instanton size and $m_q^* \approx 300 {\rm MeV}$ the constitutive quark mass in the instanton vacuum.    $m_g \approx 420 {\rm MeV}$ is the effective gluon mass in the instanton vacuum\cite{Hutter:1993sc}. In Fig.~\ref{ssafigure} (left) we  display the
results (\ref{piplus}-\ref{pizero}) as a function of the parton fraction $x_F$ for both the charged
and uncharged pions at $\sqrt{s}=19.4\,{\rm GeV}$~\cite{Skeens:1991my}. 
Fig.~\ref{ssafigure} (right) is similar to (left) except for the fact that the divergence in (\ref{firstordereq1}) is now regulated by using
a constituent gluon mass as in~\cite{Kochelev:2013zoa}.
The data in Fig.~\ref{ssafigure2} (left)  is from~\cite{Arsene:2008aa} and the data (right) is from~\cite{Adare:2013ekj}. 
In sum, the anomalous Pauli form factor
can reproduce the correct magnitude of the observed SSA in 
polarized $p_\uparrow p\rightarrow \pi X$ for reasonable vacuum parameters.  



\section{\label{sec:interactinginstanton} Spin Effects through two instantons}

\subsection{\label{subsec:doublespinresultpp} Double Spin Asymmetry in $pp$}

\begin{figure}
\includegraphics[height=40mm]{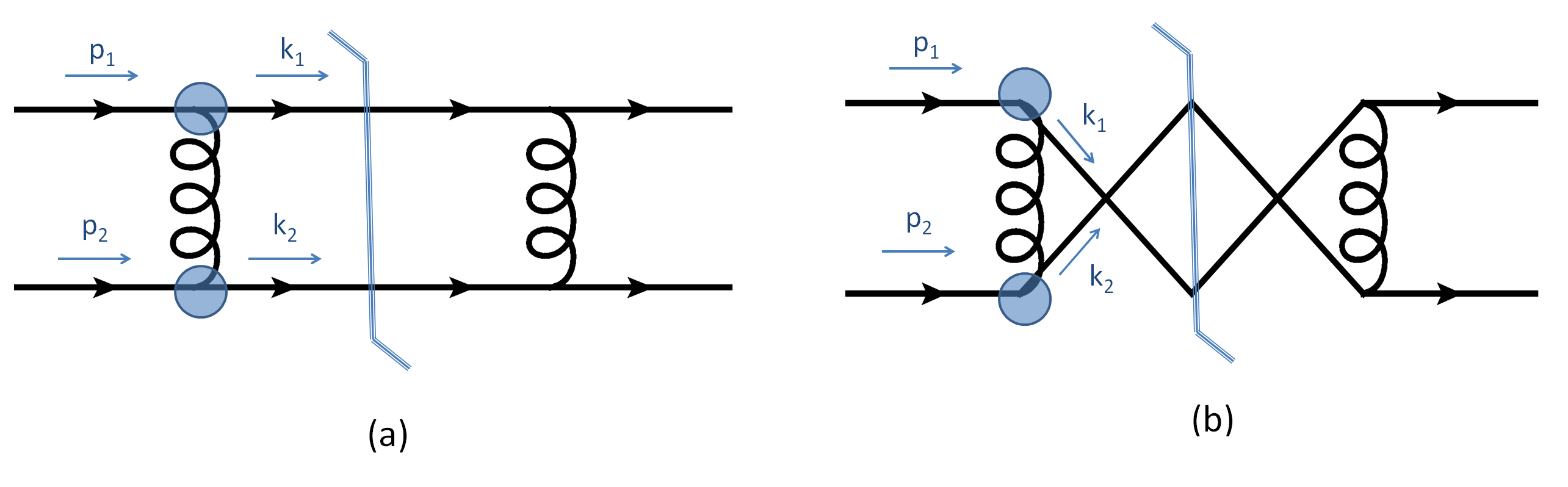}
\caption{ The valence quark in polarized proton $p_1$ exchange one gluon  with the valence quark in the polarized proton $p_2$.}
\label{dspphardgluon}
\end{figure}

 
The same Pauli form factor and vacuum parameters can be used to assess the role of the
QCD instantons on doubly polarized and semi-inclusive $p_\uparrow p_\uparrow\rightarrow \pi\pi X$
processes. The Double Spin Asymmetry (DSA) is defined as 

\be\la{doublespinasymmetry}
A_{\rm DS}= \frac{\sigma^{\uparrow \uparrow + \downarrow \downarrow} - \sigma^{\downarrow \uparrow + \uparrow \downarrow}}{\sigma^{\uparrow \uparrow + \downarrow \downarrow} + \sigma^{\downarrow \uparrow + \uparrow \downarrow}}
\ee
with the proton beam  polarized along the transverse direction. 
The valence quark from the polarized proton $P_1$ exchanges one gluon with the valence quark from the polarized proton $P_2$  as shown in~Fig.~\ref{dspphardgluon}.  At large $\sqrt{s}$,  Fig.~\ref{dspphardgluon}-(a) is dominant in forward pion production and  Fig.~\ref{dspphardgluon}-(b) is dominant in backward pion production. For Fig.~\ref{dspphardgluon}-(a), the differential cross section reads

\be 
d \sigma  \sim    \frac{g_s^4}{|p_1 - k_1|^4} \sum_{\rm color}  \tr[ M_\mu^a  \slashed{p}_1 (1 + \gamma_5 \slashed{s}_1)   \gamma_0 ( M_{\nu}^b  )^\dagger \gamma_0    \slashed{k}_1 ]    \tr[ M_\mu^a  \slashed{p}_2 (1 + \gamma_5 \slashed{s}_2)   \gamma_0 ( M_{\nu}^b  )^\dagger \gamma_0   \slashed{k}_2]   
\ee 
where $M_\mu$ is propotional to $(\bold{P}_+ +  \bold{P}_- )$ as detailed in 
~\ref{subsec:type2} of the Appendix. To second order, we approximately have
\be
\left< d^{(2)}  \sigma \right> \sim  (\cdots) \left< (\bold{P}_+ +  \bold{P}_- )^2 \right> \approx  (\cdots) \left<  \bold{P}_+ +  \bold{P}_- \right>^2
\ee
since the instanton liquid is dilute.  
The contribution to the DSA  then follows from simple algebra

\bea
\label{d2}
\left<d^{(2)}  \sigma\right>  \sim  &&   \frac{256  }{|p_1 - k_1|^4} \left(  \psi g^2_s\right)^2  [  (k_1\cdot s_1) (k_1\cdot s_2) (k_2\cdot p_1) (k_2\cdot p_2)
-(k_1\cdot p_1) (k_1\cdot s_2) (k_2\cdot p_2) (k_2\cdot s_1) \nonumber\\
&&-(k_1\cdot s_1) (k_1\cdot s_2) (k_2\cdot p_2) (p_1\cdot p_2)
+(k_1\cdot k_2) (k_1\cdot p_1) (k_2\cdot p_2) (s_1\cdot s_2)
-(k_1\cdot p_1) (k_1\cdot p_2) (k_2\cdot p_2) (s_1\cdot s_2)\nonumber\\
&&-(k_1\cdot p_1) (k_2\cdot p_1) (k_2\cdot p_2 (s_1\cdot s_2) 
+(k_1\cdot p_1) (k_2\cdot p_2) (p_1\cdot p_2) (s_1\cdot s_2)
-(k_1\cdot p_2) (k_1\cdot s_1) (k_2\cdot p_1) (k_2\cdot s_2)\nonumber\\
&&+(k_1\cdot p_1) (k_1\cdot p_2) (k_2\cdot s_1) (k_2\cdot s_2)
+(k_1\cdot k_2) (k_1\cdot s_1) (k_2\cdot s_2) (p_1\cdot p_2)
-(k_1\cdot p_1) (k_2\cdot s_1) (k_2\cdot s_2) (p_1\cdot p_2)]
\eea
after using the identity

\bea
&& \tr[(\gamma_\mu \slashed{q} -  \slashed{q} \gamma_\mu) \slashed{p} \gamma_5 \slashed{s}\gamma_\nu\slashed{k}] + \tr[\gamma_\mu \slashed{p} \gamma_5 \slashed{s}(\slashed{q}\gamma_\nu - \gamma_\nu \slashed{q})\slashed{k}] \nonumber\\
&=& \tr[(\gamma_\mu \slashed{k} +  \slashed{p} \gamma_\mu) \slashed{p} \gamma_5 \slashed{s}\gamma_\nu\slashed{k}] + \tr[\gamma_\mu \slashed{p} \gamma_5 \slashed{s}(\slashed{k}\gamma_\nu + \gamma_\nu \slashed{p}) \slashed{k}] \nonumber\\
&=& 8 i \left[ p_\mu \epsilon(\nu, k,p,s) - p_\nu \epsilon (\mu,k,p,s) + (k \cdot p) \epsilon(\mu,\nu,k,s) - (k \cdot s) \epsilon(\mu,\nu,k,p)\right]
\eea
with $q = k-p$ and $p \cdot s = 0$ because the protons are transversely polarized.
For an empirical application of (\ref{d2}) we adopt the simple kinematical set up in Eq.~\ref{kinematicssimple}. Thus

\be
\left< d^{(2)} \sigma  \right> \sim    - \frac{4  }{|p_1 - k_1|^4}  \left(  \psi g^2_s\right)^2  \tilde{s}^3 s_1^\perp s_2^\perp (1 - \cos \theta)^2 [4 + \cos (\theta - 2 \phi) + 2 \cos (2 \phi)  + \cos (\theta + 2 \phi)]
\ee
After adding the contribution of  Fig.~\ref{dspphardgluon}-(a) and Fig.~\ref{dspphardgluon}-(b), 
and averaging over the transverse direction $\phi$, we finally  obtain
\be
\frac{d^{(2)}  \sigma}{d^{(0)}  \sigma} \sim  - 4  s_1^\perp s_2^\perp   \left(  \frac{ \pi^4  n_I \rho^4_c}{m_q^* g_s^2} \right)^2    \frac{ F_g^2 [\rho \sqrt{\frac{\tilde{s}(1 - \cos \theta)}{2}} ]  \tilde{s}   + F_g^2 [\rho \sqrt{\frac{\tilde{s}(1 + \cos \theta)}{2}} ]  \tilde{s}    }{\frac{5 + 2 \cos \theta +   \cos^2 \theta}{ (1 - \cos \theta)^2}  + \frac{5 - 2 \cos \theta +   \cos^2 \theta}{ (1 +\cos \theta)^2} }
\ee

Our DSA results can now be compared to future experiments at collider energies.
Specifically, our DSA for dijet productions are

\be\la{doublepiplus}
A_{\pi^+ \pi^+} = - \frac{1}{8}  \frac{\Delta_s u(x_1  ,Q^2)}{u(x_1  , Q^2)}   \frac{\Delta_s u(x_2  ,Q^2)}{u(x_2 , Q^2)} \left(  \frac{ \pi^3  n_I \rho^4_c}{m_q^* \alpha_s }   \right)^2      \frac{ F_g^2 [\rho \sqrt{\frac{\tilde{s}(1 - \cos \theta)}{2}} ]  \tilde{s}     + F_g^2 [\rho \sqrt{\frac{\tilde{s}(1 + \cos \theta)}{2}} ]  \tilde{s}  }{ (5 + 10 \cos^2 \theta + \cos^4 \theta) \csc^4 \theta }
\ee

\be\la{doublepiminus}
A_{\pi^- \pi^-} =  - \frac{1}{8} \frac{\Delta_s d(x_1  ,Q^2)}{d(x_1  , Q^2)}   \frac{\Delta_s d(x_2  ,Q^2)}{d(x_2 , Q^2)}  \left(  \frac{ \pi^3  n_I \rho^4_c}{m_q^* \alpha_s }   \right)^2      \frac{ F_g^2 [\rho \sqrt{\frac{\tilde{s}(1 - \cos \theta)}{2}} ]  \tilde{s}     + F_g^2 [\rho \sqrt{\frac{\tilde{s}(1 + \cos \theta)}{2}} ]  \tilde{s}  }{ (5 + 10 \cos^2 \theta + \cos^4 \theta) \csc^4 \theta }
\ee

\bea\la{doublepiplusminus}
A_{\pi^+ \pi^-} =&&  -  \frac{1}{8} \frac{\Delta_s u(x_1  ,Q^2) \Delta_s d(x_2  ,Q^2)+\Delta_s d(x_1  ,Q^2) \Delta_s u(x_2  ,Q^2)}{u(x_1  , Q^2)d(x_2 , Q^2)+d(x_1  , Q^2)u(x_2 , Q^2)}   \left(  \frac{ \pi^3  n_I \rho^4_c}{m_q^* \alpha_s }   \right)^2   \nonumber\\
&& \times  \frac{ F_g^2 [\rho \sqrt{\frac{\tilde{s}(1 - \cos \theta)}{2}} ]  \tilde{s}     + F_g^2 [\rho \sqrt{\frac{\tilde{s}(1 + \cos \theta)}{2}} ]  \tilde{s}  }{ (5 + 10 \cos^2 \theta + \cos^4 \theta) \csc^4 \theta }
\eea
To compare our calculations with the experimental results, we use the same kinematics in Fig.~\ref{ssafigure2}: $\sqrt{s} = 62.4 {\rm GeV}$ and $\eta = 3.5$. The 
value of $\alpha_s$ is from~\cite{Beringer:1900zz}.   Our predictions for charged di-jet production
in semi-inclusive DSA are displayed in  Fig.~\ref{doublegraphs}.

\begin{figure}[!htb]
\includegraphics[height=50mm]{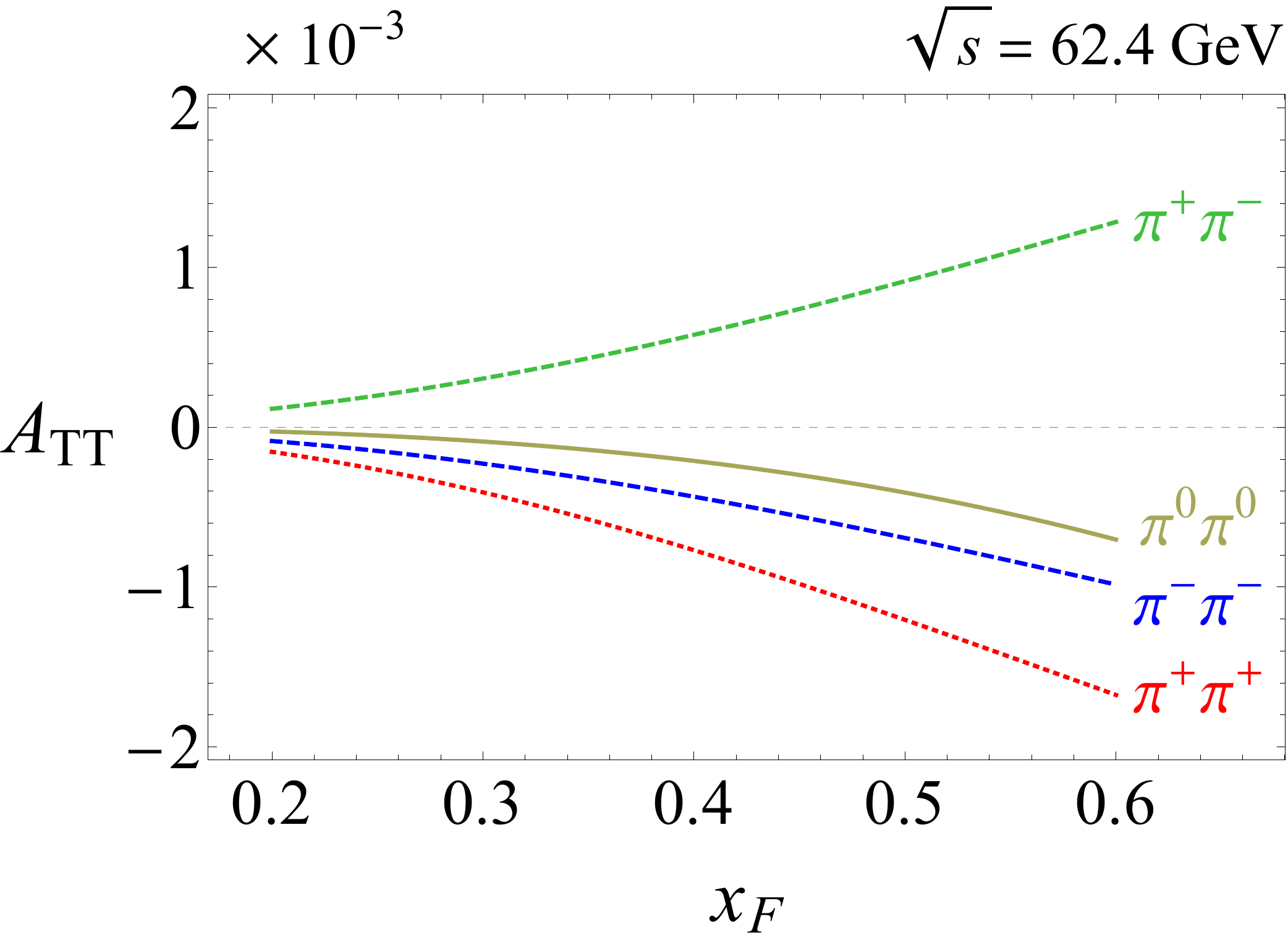}
\caption{\la{doublegraphs} Predictions for charged di-jet production
in semi-inclusive DSA.}
\end{figure}

\section{\label{subsec:poddeffectresult} $P$-odd effects through Instanton Fluctuations}

 It is commonly accepted that in a typical non-central $AuAu$ collision at RHIC as illustrated 
in Fig.~\ref{auaucollision} (left), the flying fragments create a large magnetic field that strongly 
polarizes the wounded  or participant nucleons in the final state.  The magnetic field is typically 
$eB/m_\pi^2\approx 1$ at RHIC and $eB/m_\pi^2\approx 15$ at the LHC and argued to last
 for about 1-3 ${\rm fm}/c$~\cite{Skokov:2009qp}.
We recall that in these units  $m_\pi^2\approx 10^{18}\,{\rm Gauss}$ which is substantial.
As a result, large ${\cal P}$-odd pion  azimuthal charge correlations were predicted to take place
in peripheral heavy ion collisions~\cite{Kharzeev:1998kz,Kharzeev:2004ey,Kharzeev:2007jp,Fukushima:2008xe}.

In this section, we would review the analysis in~\cite{Qian:2012eq} and show that a large contribution to the ${\cal P}$-odd pion azimuthal 
charge correlations may stem from the prompt part of the collision as 
each of the incoming nucleus polarizes strongly the participating nucleons from its partner nucleus
during the collision process as illustrated in~Fig.~\ref{auaucollision} (right). The magnetic field is strong but
short lived in the initial state, lasting for about $1/2\,{\rm fm}/c$ for a typical heavy ion collision
at current collider energies. Polarized proton on proton scattering can exhibit large chirality flip
effects through instanton and anti-instanton fluctuations as we now show.

\subsection{$\mathcal{P}$-odd Effects in the Instanton Vacuum \la{sec:podd}}

\begin{figure}[!htb]
\minipage{0.45\textwidth}
\includegraphics[height=50mm]{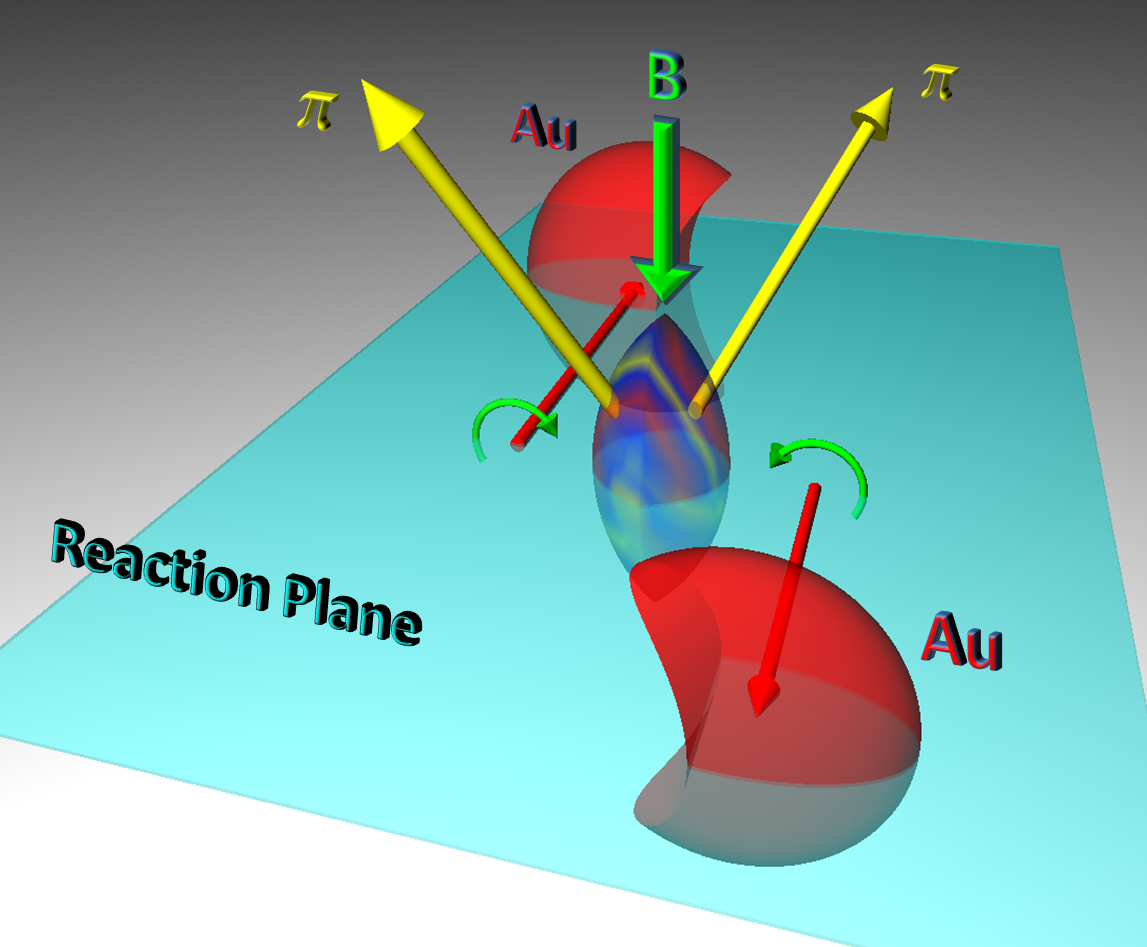}
\endminipage\hfill
\minipage{0.45\textwidth}
\includegraphics[height=50mm]{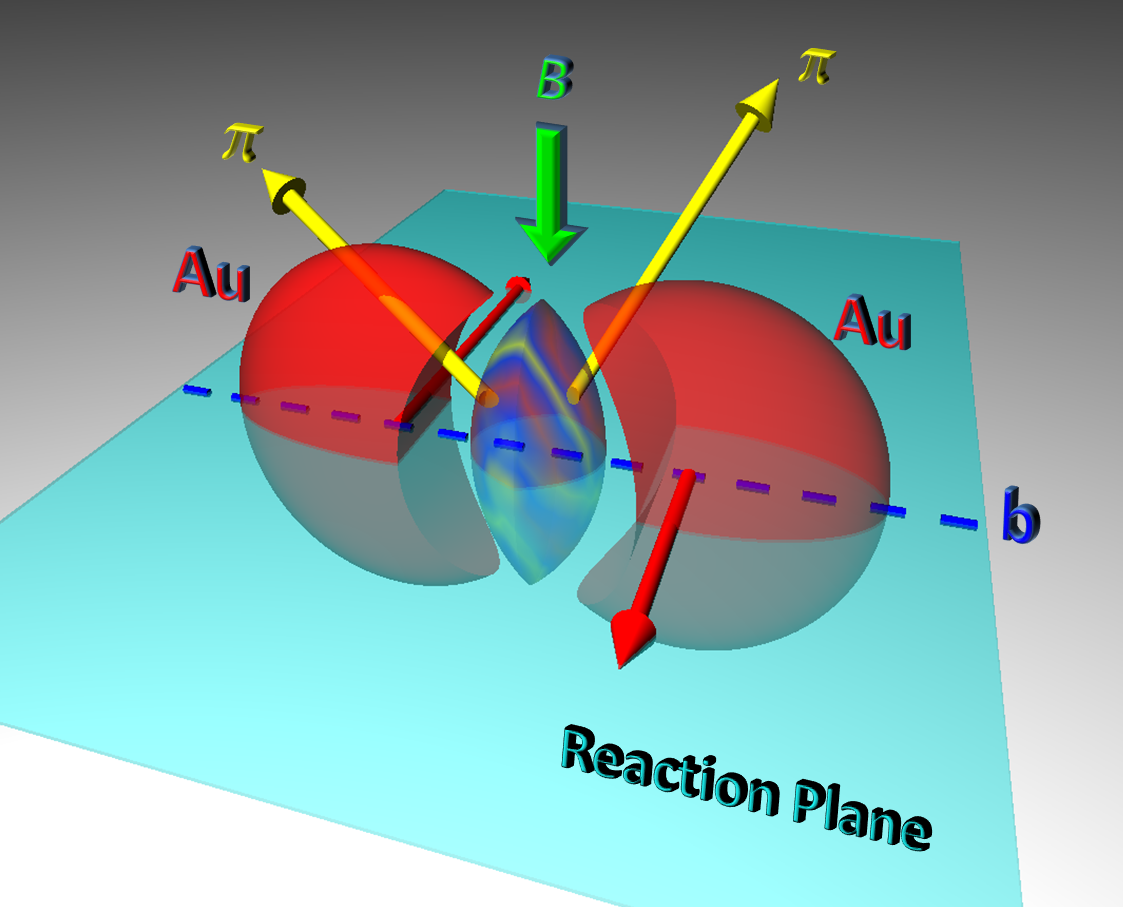}
\endminipage
\caption{2-pion correlations in $AuAu$  after the collision. \la{auaucollision} }
\end{figure}

\begin{figure}
\includegraphics[height=32mm]{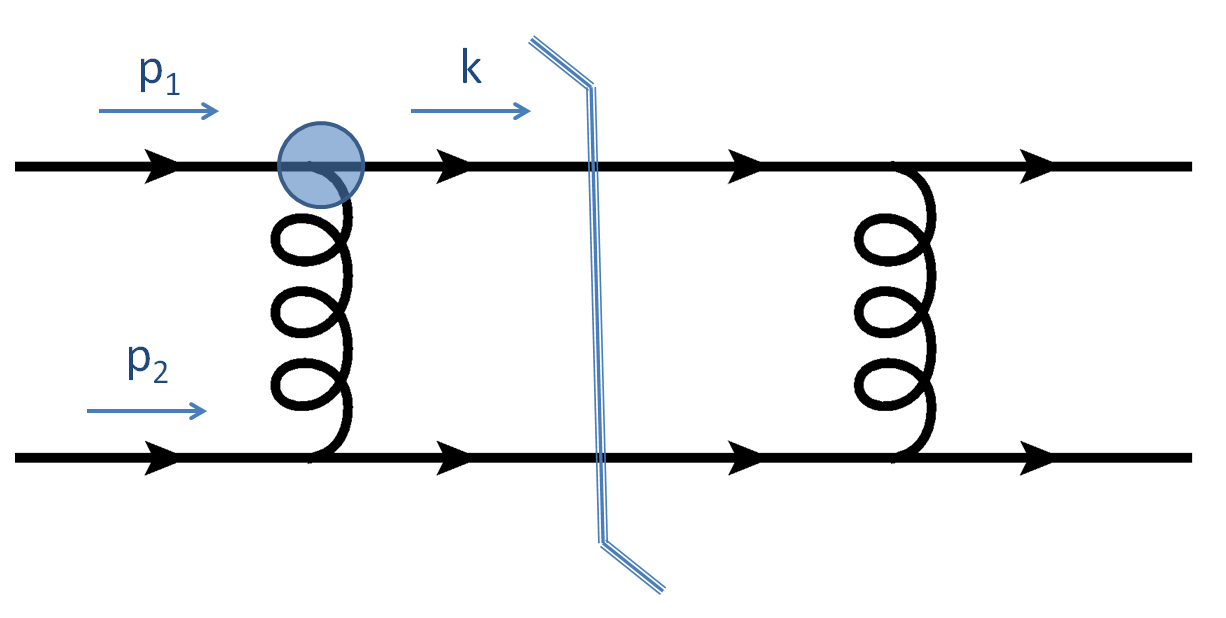}
\caption{ Gluon Exchange. The blob is an instanton or anti-instanton insertion. See text.}
\label{gluonexchange}
\end{figure}

Consider the typical parton-parton scattering amplitude of Fig.~\ref{gluonexchange} with 2-gluon exchanges.
In each collision, the colliding  ``parton''  $p_i$ has spin $s_i$, and thus $u(p_i) \bar{u}(p_i) = \frac{1}{2} \slashed{p_i}(1 + \gamma_5 \slashed{s}_i)$. The parton $p_1$ from the A-nucleus encounters an instanton or anti-instanton 
as depicted by the gluonic form-factor.  Rewrite the \eqref{VERTEX} 

\be  
M_\mu^a = t^a  \left[  \gamma_\mu-   \bold{P}_+ \gamma_+    \sigma_{\mu\nu} q^\nu    \Psi     -   { \bold{P}}_- \gamma_-     \sigma_{\mu\nu} q^\nu  \Psi   \right]
\ee
with $\bold{P}_+ = 1$ stands for an instanton insertion and ${\bold{P}}_-=1$  for an  anti-instanton insertion. 
 In establishing (\ref{VERTEX}),  the instanton and anti-instanton zero modes are assumed to be 
undistorted by the prompt external magnetic field. Specifically, the chromo-magnetic field $B_G$
is much stronger than the electro-magnetic field $B$, i.e. $|g_sB_G|\gg |eB|\approx $ or
$m_\pi^2\rho_c^2\approx 0.004\ll 1$. The deformation of the instanton zero-modes by a strong magnetic field
have been discussed in~\cite{Basar:2011by}. They will not be considered here.

In terms of (\ref{VERTEX}), the contribution of Fig.\ref{gluonexchange} to the differential cross section is

\be 
d \sigma  \sim   \frac{g_s^4}{ |p_1 - k|^4}    \tr[ M_\mu^a  \slashed{p}_1 (1 + \gamma_5 \slashed{s}_1 )   \gamma_0 (M_\nu^b)^\dagger   \gamma_0 \slashed{k}  ]  \tr[\gamma^\mu t_a \slashed{p}_2 (1+\gamma_5 \slashed{s}_2)\gamma^\nu t_b \slashed{k}^\prime]
\ee  
which can be decomposed into $d\sigma\approx d\sigma^{(0)}+d\sigma^{(1)}$  in the dilute instanton liquid.
The zeroth order contribution is

\be \la{zeroorderdiff}
d^{(0)} \sigma  \sim    \frac{64 g_s^4}{  |p_1 - k|^4}  \left[   2 (k  \cdot p_2) (p_1 \cdot p_2) +  (k  \cdot p_1) (p_1 \cdot p_2 - k\cdot p_2)    \right]
\ee 
where we used $k^\prime = p_1 + p_2 - k$. The first order contribution is 

\be 
d^{(1)} \sigma  \sim   \frac{64  g_s^4  }{ |p_1 - k|^4}     \left[  (p_1 \cdot p_2)^2 + (k \cdot p_2) (p_1 \cdot p_2)    \right] (k \cdot s_1) \Psi  \left(   \bold{P}_+ -   { \bold{P} }_-    \right) 
\ee 
after using $p_1 \cdot s_1 =  0$ and $p_1^2 = k^2 = 0$. Converting to standard parton kinematics with
$p_1 \rightarrow x_1 P_1$, $p_2 \rightarrow x_2 P_2$ and $k \rightarrow {K}/{z}$, we
obtain for the ratio of the $\mathcal{P}$-odd to $\mathcal{P}$-even contributions in the differential cross section

\be 
\frac{d^{(1)} \sigma }{d^{(0)} \sigma }  =   \frac{ x_1   (P_1 \cdot P_2)^2 + \frac{1}{z} (K \cdot P_2) (P_1 \cdot P_2)   }{ 2    (K  \cdot P_2) (P_1 \cdot P_2) +   (K \cdot P_1) ( \frac{x_1}{x2} P_1 \cdot P_2 -\frac{ K \cdot P_2}{z x2 } )}    (K \cdot s_1) \Psi  \left(   \bold{P}_+ -   { \bold{P} }_-    \right) 
\ee

Now consider the kinematics appropriate for the collision set up in Fig.~\ref{auaucollision},
\bea\la{kinematicssimple}
P_{1/2} &=&  \frac{\sqrt{s}}{2} (1, 0, 0, \pm 1) \nonumber\\
K &=& (E ,  K_\perp \cos \Delta\phi, K_\perp \sin \Delta\phi ,  \frac{\sqrt{s}}{2} x_F ) \nonumber\\
s_1 &=&  (0,0, s^\perp_1 , 0)
\eea
where $K_\perp$  and $E ^2= {K_\perp^2 + s x_F^2/4 + m_\pi^2 }$ are  the transverse momentum and total squared energy of the outgoing pion respectively. $x_F$ is the pion longitudinal momentum fraction. Thus

\be \la{crosssection}
\lim_{s \rightarrow \infty}  \frac{ d^{(1)} \sigma }{d^{(0)} \sigma} =     (\sin \Delta\phi)  s^\perp_1  \frac{x_F + x_1 z}{  x_F z}  \frac{K_\perp}{m_q^*}  \frac{    \pi^3   (n_I \rho^4_c)   }{ 8 \alpha_s}  F_g \left(\rho \sqrt{ \frac{x_1 }{x_F z} (K_\perp^2 + m_\pi^2)}\right) \left(   { \bold{P} }_-  - \bold{P}_+   \right)  
\ee 
We note that Eq.~\ref{crosssection} vanishes after averaging over the instanton liquid background which is $\mathcal{P}$-even

\be
\left<  \frac{d^{(1)} \sigma }{d^{(0)} \sigma }  \right> = 0
\ee
since on average $\left< \bold{ Q } \right> = \left<  \bold{P}_+  -   { \bold{P} }_-   \right> = 0$.

\subsection{$\mathcal{P}$-odd effects  in AA Collisions\la{sec:aacollision}}

Now consider hard $pp$ collisions in peripheral $AA$ collisions as illustrated inFig.~\ref{auaucollision} (right). The Magnetic field is strong enough at the collision to partially polarize the colliding protons. Say $c\%$ of the wounded protons from a given nucleus get polarized by the partner colliding nucleus.
For simplicity,  we set $s_\perp u(x,Q^2) = c\%  \Delta_s u (x,Q^2)$ and $s_\perp d(x,Q^2) =c\% \Delta_s d (x,Q^2)$, with $\Delta_s u (x,Q^2)$ and $\Delta_s d (x,Q^2)$ as the spin polarized distribution functions of the valence up-quarks and valence down-quarks in the proton respectively. We also assume that the outgoing $u$ quark turns to $\pi^+$ and that the outgoing $d$ quark  turns to $\pi^-$. With this in mind, we may rewrite the ratio of differential contributions
in (\ref{crosssection}) following~\cite{Abelev:2009ac,Abelev:2009ad,collaboration:2011sma,Selyuzhenkov:2012py} as

\be
\frac{d {\bf N}}{ d \phi_{\alpha}} \sim 1 -  2 a_{\alpha} \sin (\phi - \Psi_{RP})
\ee
with $\alpha=\pm $ or 
\be 
a_+  =    \frac{\Delta_s u(x  ,Q^2)}{u(x  , Q^2)}    \Upsilon  \bold{ Q }  \ \ \ \ \ \ \ \ 
a_-  =     \frac{\Delta_s d(x  ,Q^2)}{d(x  , Q^2)}  \Upsilon  \bold{ Q } 
\ee 
and
\be
\Upsilon \equiv \frac{ x_F   +   x     z   }{    x_F    z   }  \frac{K_\perp}{m_q^*}  \frac{    \pi^3   (n_I \rho^4_c)   }{ 16 \alpha_s}  F_g \left(\rho \sqrt{ \frac{  x    }{  x_F  z } (K_\perp^2 + m_\pi^2)}\right)
\ee
While on average $\left<a_\alpha \right>=0$ since $\left<{\bf Q}\right>_V=0$, in general $\left< a_{\alpha} a_{\beta} \right> \neq 0$
for the 2-particle correlations.
Explicitly

\bea
- \left< a_{\pi^+}  a_{\pi^-}  \right> &=& - \left( \frac{\Delta_s u(x  ,Q^2)}{u(x  , Q^2)}  \frac{\Delta_s d(x  ,Q^2)}{d(x  , Q^2)} \right)   \Upsilon^2   \left<  \bold{Q}^2 \right>_V   \nonumber\\ 
-  \left< a_{\pi^+}  a_{\pi^+}  \right>&=& - \left( \frac{\Delta_s u(x  ,Q^2)}{u(x  , Q^2)} \right)^2   \Upsilon^2   \left<  \bold{Q}^2 \right>_V    \nonumber\\
- \left< a_{\pi^-}  a_{\pi^-}  \right> &=& - \left( \frac{\Delta_s d(x  ,Q^2)}{d(x  , Q^2)} \right)^2   \Upsilon^2   \left<  \bold{Q}^2 \right>_V  
\eea
 For reasonable values of $\left< x \right>$, $\left< a_{\pi^+}  a_{\pi^+}  \right> \sim \left< a_{\pi^-}  a_{\pi^-}  \right> \sim -\left< a_{\pi^+}  a_{\pi^-}  \right>$ as expected~\cite{Abelev:2009ac,Abelev:2009ad,collaboration:2011sma,Selyuzhenkov:2012py}.

A more quantitative comparison to the reported data in ~\cite{Abelev:2009ac, Selyuzhenkov:2012py} can be carried out by estimating the fluctuations
of the topological charge ${\bf Q}$ in the prompt collision 4-volume $V\approx (\tau^2/2) \Delta\eta V_\perp(b)$. 
In the latter, $\tau\approx$  1/2
${\rm fm}/c$ is the prompt proper time over which the induced magnetic field is active, $\Delta\eta$ is the interval in
pseudo-rapidity and $V_\perp(b)$ the transverse collision area for fixed impact parameter $b$. Through simple
geometry

\be
V_\perp(b)=2R^2\left({\rm arccos}\left(\frac b{2R}\right)-\frac b{2R}\sqrt{1-\left(\frac b{2R}\right)^2}\right)
\ee
where $R$ is the radius of two identically colliding nuclei. ${\bf Q}^2$ involves a pair ${\bf P}, {{\bf P}^\prime}$ of instanton-antiinstanton.
Specifically,

\be
\label{PAIR1f}
\left<{\bf Q}^2\right>_V=\left<({\bf P}_+-{\bf P}_-)({\bf P}_+^\prime-{{\bf P}_-^\prime})\right>_V
\ee
If we denote by $N_\pm$ the number of instantons and antinstantons in $V$, 
with $N=N_++N_-$ their total number, then in the instanton vacuum the pair correlation follows from

\be
\left<{\bf Q}^2\right>_V  \equiv  \left<\left(\frac{N_+-N_-}{N_++N_-}\right)^2\right>_V  \approx \frac{\left<(N_+-N_-)^2\right>_V}{\left<(N_++N_-)^2\right>_V}
\ee
Assuming $N_\pm$ to be large in $V$ it follows that~\cite{Diakonov:1983hh}

\be 
\label{PAIR1}
\left<{\bf Q}^2\right>_V  \approx  \frac{\left<N\right>_V}{\left<N\right>_V
(\left<N\right>_V+4/{\bf b})}
\ee 
The deviation from the Poissonian distribution in the variance of the number average reflects on the
QCD trace anomaly in the instanton vacuum  or $\left<N^2\right>_V-\left<N\right>^2_V=4/{\bf b} \left<N\right>_V$
and vanishes in the large $N_c$ limit~\cite{Diakonov:1983hh}. Here ${\bf b}=11N_c/3$  is the coefficient of the 1-loop beta function $\beta(\rho_c)\approx {\bf b}/{\rm ln}(\Lambda\rho_c)$ (quenched).  Thus

\be
\left<{\bf Q}^2\right>_V\approx \frac 1{n_I(\tau^2\Delta\eta V_\perp(b)/2)+4/{\bf b}}
\ee
where we have used that the mean $\left<N\right>_V=n_IV$ in the volume $V$. 
The topological fluctuations are suppressed by the  collision 4-volume. Note that we have ignored the role
of the temperature on the the topological fluctuations in peripheral collisions. Temperature will cause these topological
fluctuations to deplete and vanish at the chiral transition point following the instanton-anti-instanton pairing~\cite{Janik:1999ps}.  So our results will be considered as upper-bounds.

\begin{figure}[!htb]
\minipage{0.33333333333 \textwidth}
\includegraphics[height=45mm]{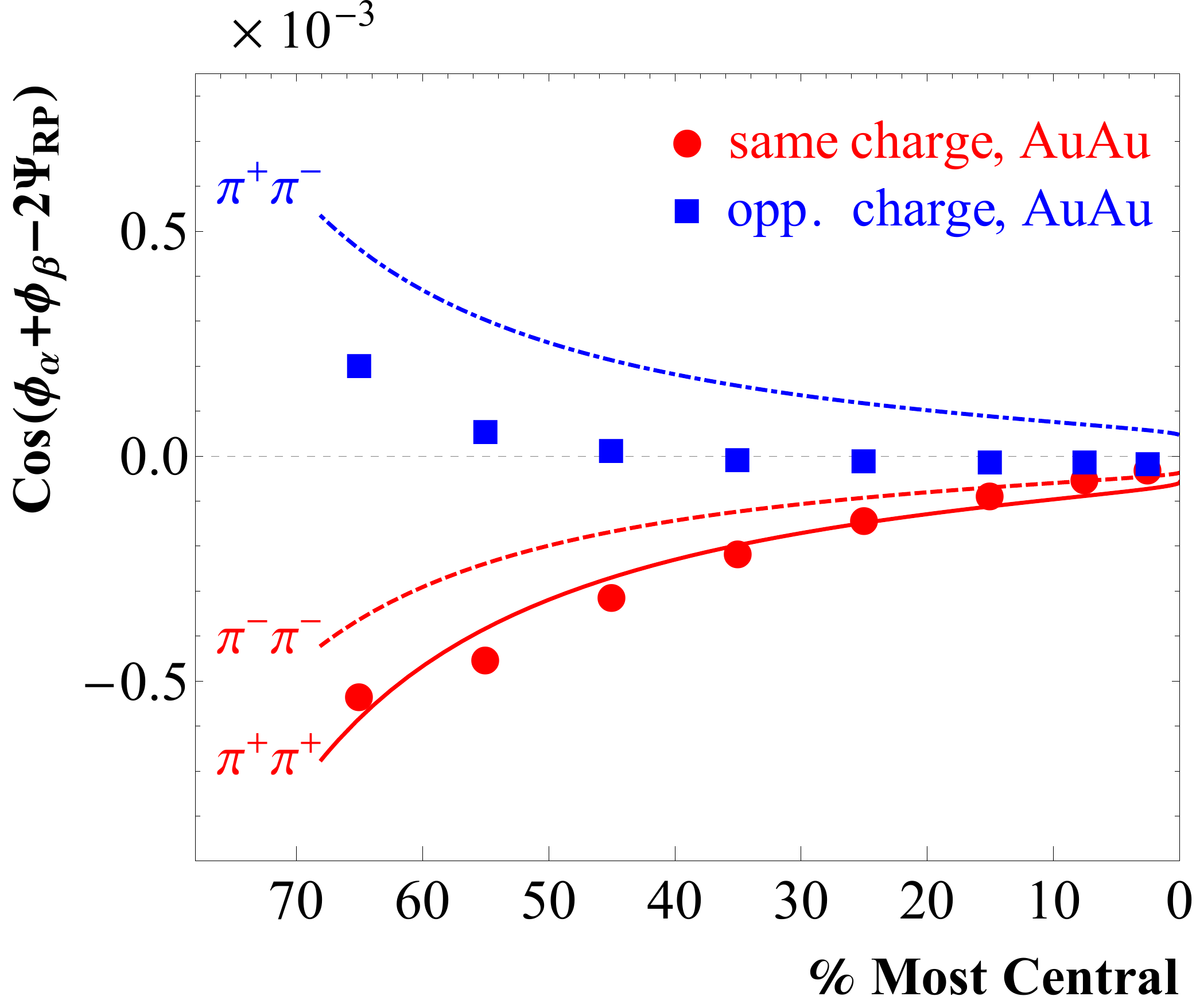}
\endminipage\hfill
\minipage{0.33333333333 \textwidth}
\includegraphics[height=45mm]{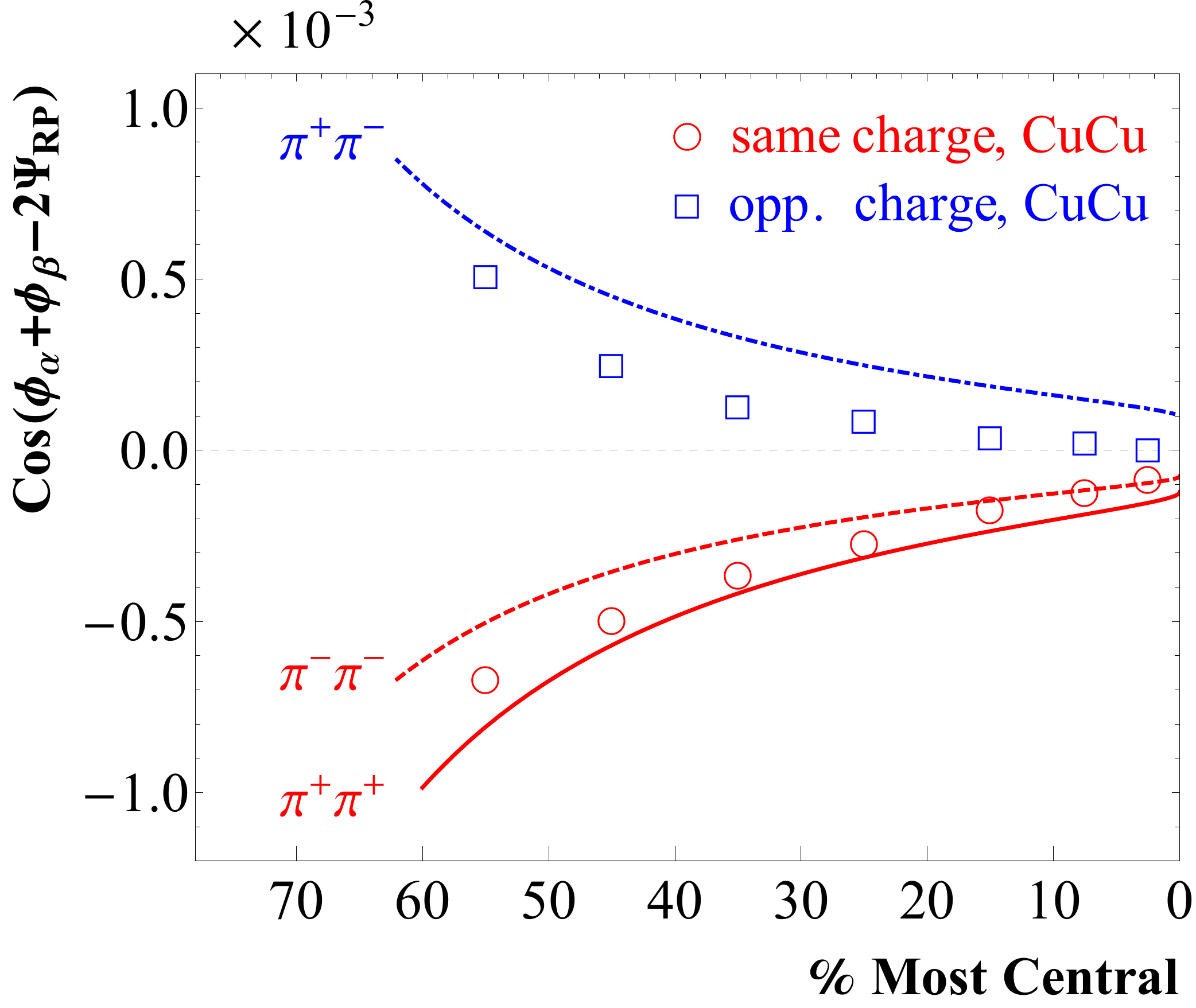}
\endminipage
\minipage{0.33333333333 \textwidth}
\includegraphics[height=45mm]{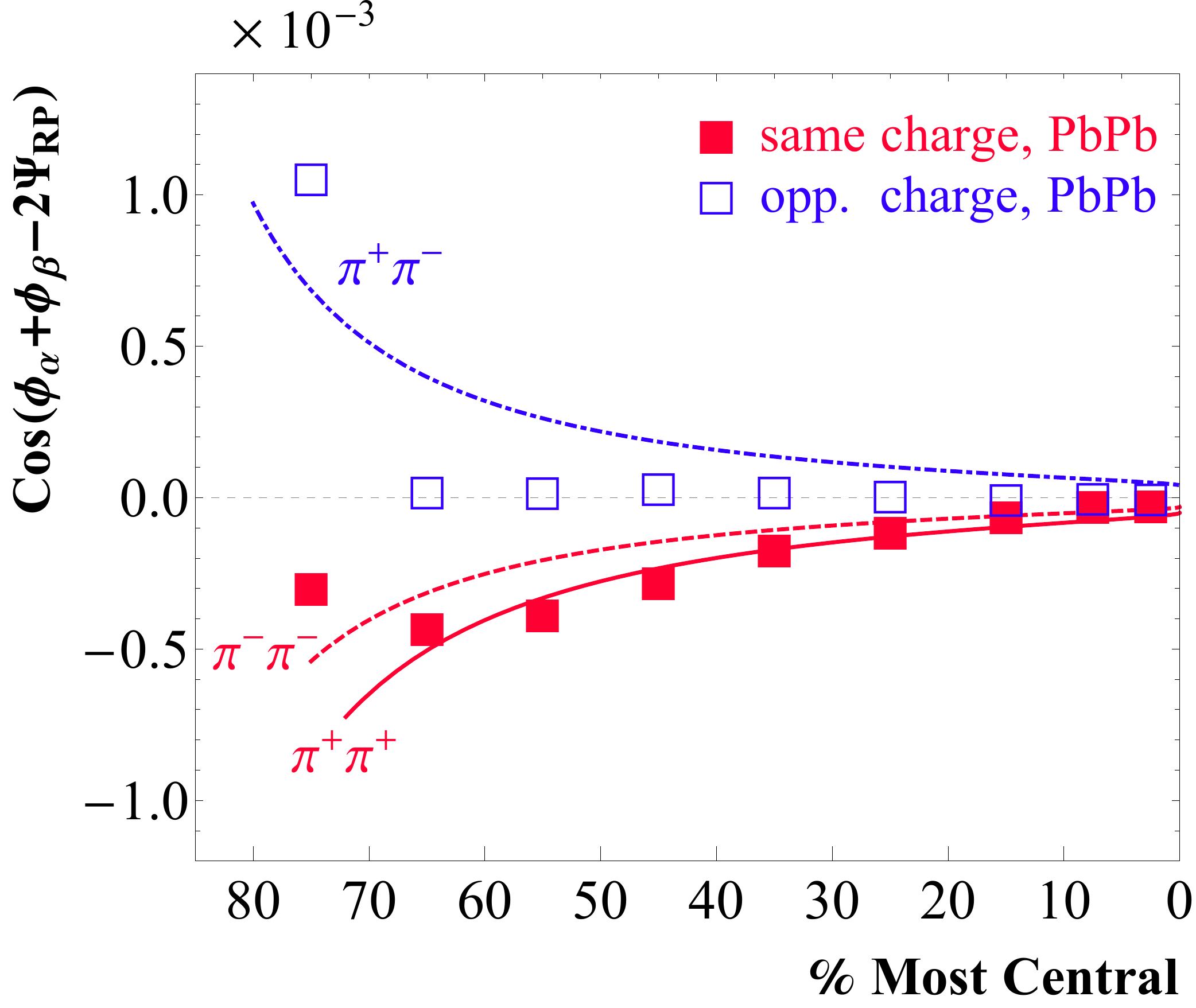}
\endminipage
  \caption{    Pion azimuthal charge correlations versus the data~\cite{Abelev:2009ac} from STAR  at $\sqrt{s}= 200 {\rm GeV}$ (left and middle) and the data from ALICE~\cite{Selyuzhenkov:2012py}  at $\sqrt{s}=2.76 {\rm TeV}$ (right).    }\label{poddacp}
\end{figure}

For simplicity, we will set  $\left< x\right> \approx 0.01$ for each parton and $\left<z \right> \approx 0.5$. 
The measured multiplicity spectra in~\cite{Khandai:2011cf} at different centralities suggest 
$m_\pi<\left<K_\perp\right> <3 m_\pi$. We will set $\left<K_\perp\right> =2 m_\pi$ in our analysis.
We will assume a moderate polarization or $c\% = 15 \%$ in the collision volume for a general discussion.  
We fix $\tau = 1/2 \,{\rm fm}$ to be the maximum duration of the magnetic field polarization, and set
the pseudo-rapidity interval approximately $( -1, 1)$ for both STAR~\cite{Abelev:2009ac}  and ALICE~\cite{Selyuzhenkov:2012py}.
The radius of the colliding nuclei 
will be set to $R=1 {\rm fm} \times \sqrt[3]{A} $ where $A$ is the atomic number.
 The centrality is approximated as 
$n\% = b^2/(2 R)^2$~\cite{Aguiar:2004uq}. Our results are displayed in Fig.~\ref{poddacp} (left) for  $AuAu$ and  Fig.~\ref{poddacp} (middle) for $CuCu$ 
collisions  at $\sqrt{s}=200 {\rm GeV}$ (STAR), and  in Fig.~\ref{poddacp} (right) for $PbPb$ collisions  at $\sqrt{s} = 2.76 {\rm TeV}$
(ALICE).  We recall that~\cite{Voloshin:2004vk}

\be
\left< \cos (\phi_\alpha + \phi_\beta - 2 \Psi_{\rm RP}) \right> \equiv - \left< a_{\alpha} a_{\beta} \right> 
\ee
For the like-charges the results compare favorably with the  data. For the unlike charges they overshoot the  data 
especially for the heavier ion. Our results show a difference between $\pi^+\pi^+$ and $\pi^-\pi^-$ as we only retained
the protons in our analysis. The inclusion of the neutrons would result into the same charge correlations for 
$\pi^+\pi^+$ and $\pi^-\pi^-$ by isospin symmetry.

\section{Conclusions and Prospects}

Instantons and anti-instantons provide the key building blocks of the instanton liquid model.
The latter offers a detailed framework for understanding aspects of the spontaneous
breaking of chiral symmetry and the resolution of the U(1) problem.  Key to this is the 
appearance of light quark zero modes of fixed chirality and their de-localization through
the formation of an interacting liquid. Some aspects of this model are supported by 
lattice simulations upon cooling~\cite{Chu:1992mn,Chu:1993cn,Chu:1994vi}.

In light of the many phenomenological successes of the instanton liquid model, it is natural 
to ask about the role of instantons in scattering processes, in particular on spin physics. An essential aspects
of the light quark zero modes is the emergence of large constituent masses and  (chromo) magnetic moments. 
Also instantons and anti-instantons correlate strongly the spin with color leading to sizable contributions in spin
polarized processes  involving light quarks. 

This review gives a brief summary of recent advances in the emerging field of spin physics where the
induced effects by instantons and anti-instantons in a semi-classical analysis, are sizable in comparison to
those usually parametrized using perturbation theory. We stress that the effects we have reported 
both  in polarized electron-proton or proton-proton  semi-inclusive scattering,
 rely solely on the instanton liquid parameters in the vacuum without additional changes.
 The effects are large and comparable in size with those reported experimentally.

 This review also shows that the large spin effects induced by instantons and anti-instantons in polarized experiments
 may also be present in peripheral $AA$ collisions where  a prompt and large magnetic field can induce a prompt and
 large polarization although on a short time scale.  A simple analysis of the correlated fluctuations between target and projectile
 protons shows that the effects is of the same magnitude and sign are those reported in the peripheral charged pion 
 azimuthal correlations at collider energies. Again it is important to stress that only the fluctuations expected from 
 instanton vacuum configurations were used.

 This review is by no means exhaustive as many new effects can be explored using this framework.
 One important shortcoming of the instanton liquid model is the lack of confinement as described by 
 an ordering of the eigenvalues of the Polyakov line at low temperature. Some important amendments
 to the instanton liquid model have been proposed, suggesting that instantons and anti-instantons split
 into dyons in the confined phase~\cite{Diakonov:2007nv}. It was recently shown that the key chiral effects and
 U(1) effects in the standard instanton liquid model are about similar to those emerging from the
 new instanton-dyon liquid model~\cite{Shuryak:2012aa,Shuryak:2014gja,Larsen:2014yya,Larsen:2015vaa,Larsen:2015tso,Liu:2015ufa,Liu:2015jsa}. It would be important to revisit the spin effects in this context.

\section{\label{sec:acknowledgements}acknowledgements}

This work was supported in parts by the US-DOE grant DE-FG-88ER40388.

\section{\label{app:instantoninsertion} Appendix:  Effective vertex in instanton background}

\subsection{\label{subsec:type1} Photon vertex}

In this Appendix, we review the derivation of (\ref{dipm1}) in~\cite{Qian:2011ya}, corresponding to the nonperturbative insertion $\tilde{M}_\mu^{(1)}$ for photon exchange in the single instanton background. Similar calculations can also be found in~\cite{Moch:1996bs, Ostrovsky:2004pd}. According to~\cite{Brown:1977eb,Moch:1996bs,Faccioli:2001ug,Ostrovsky:2004pd}, the zero mode quark propagator in the single instanton background after Fourier transformation with respect to the incoming momentum $p$ is

\be\la{simplify1}
{S_0 (x,p)_{\dot{\beta}}^{\ \ j }}_{i \delta}= \frac{2 \rho^2}{ \lambda} \frac{x^l (\overline{\sigma}_l)_{\dot{\beta} \gamma} \varepsilon^{\gamma j} \varepsilon_{i \delta}   }{(x^2 + \rho^2)^\frac{3}{2} |x|}
\ee
Note the chirality of the zero mode flips as $\left|L\right>\left<R\right|$ as depicted in Fig.~\ref{flip}. The incoming quark is left-handed and has momentum $p$ (on-shell). $\rho$ is the size of instanton and $\lambda$ is the mean virtuality. $\beta$ and $\delta$ are spatial indices, while $j$ and $i$ are color indices. In Euclidean space, $\sigma_\mu = (\vec{\sigma}, i I)$, $\bar{\sigma}_\mu = (\vec{\sigma}, i I)$ and $\epsilon^{01} =-\epsilon^{10}= -\epsilon_{01} = \epsilon_{10}$~\cite{Vandoren:2008xg}.
The right-handed non-zero mode quark propagator in the single instanton after Fourier transformation with respect to the outgoing momentum $k$ is~\cite{Brown:1977eb,Moch:1996bs,Ostrovsky:2004pd}

\be\la{simplify2}
{S_{nz} (k, x)^{\beta i}}_{j \alpha} = - \delta_\alpha^\beta  \left( \delta^i_j + \frac{\rho^2}{x^2} \frac{(\sigma_\rho \overline{\sigma}_r)^i_{\ \ j} k^\rho x^r}{2 k \cdot x} \left(1 - e^{- i k \cdot x} \right) \right)  \frac{|x|}{\sqrt{x^2 + \rho^2}} e^{i k \cdot x} 
\ee

\begin{figure}
\includegraphics[height=40mm]{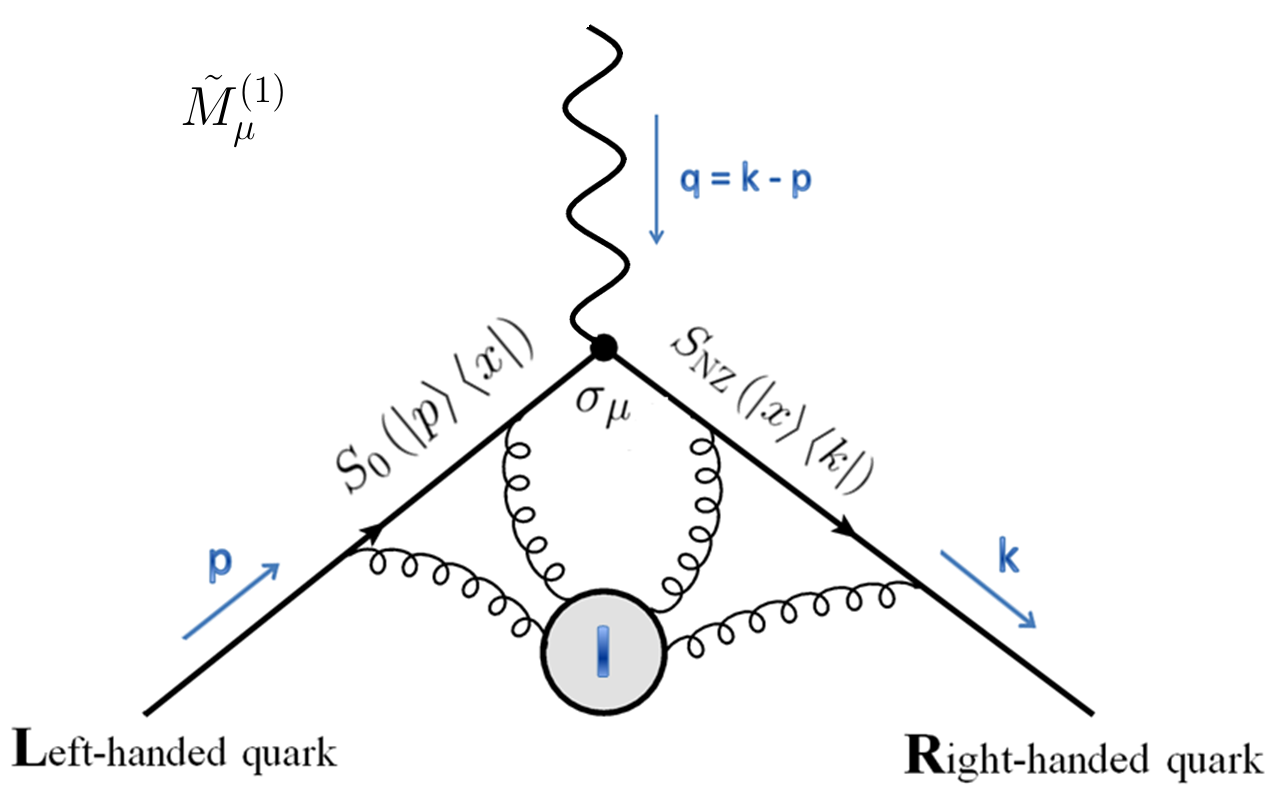}
\caption{\la{flip} The incoming left-handed quark with momentum $p$ meets one instanton and flips its chirality. The outgoing right-handed quark carries momentum $k$. The momentum of the  photon is $q = p - k$. $S_0$ and $S_{nz}$ stand for the zero-mode quark propagator and the non-zero mode quark propagator in the single instanton background respectively.  }
\end{figure}

Consider the process depicted in Fig.~\ref{flip}: the incoming left-handed quark meets one instanton and flips its chirality (zero-mode), 
then  exchanges one photon, and finally becomes an outgoing right-handed  quark. As a result, the nonperturbative insertion $M_\mu^{(1)}$ reads

\be\la{beforesimplify}
{\left(\tilde{M}_\mu^{(1)}\right)^{\beta i}}_{i^\prime \delta} = \int d^4 x \quad e^{- i q \cdot x} \, {S_{nz} (k, x)^{\beta i}}_{j \alpha} \, \sigma_\mu^{\alpha \dot{\beta}} \, {S_0 (x,p)_{\dot{\beta}}^{\ \ j }}_{i^\prime \delta}
\ee
All the other parts of the diagram are trivial in color, therefore we take the trace of color indices $i$ and $i^\prime$. 
To further simplify the result, we need the following formula~\cite{Vandoren:2008xg}

\be
  \delta_\alpha^\beta  \delta^i_j  (\sigma_\mu)^{\alpha \dot{\beta}}  (\overline{\sigma}_l)_{\dot{\beta} \gamma} \varepsilon^{\gamma j} \varepsilon_{i \delta}  = {(\sigma_\mu \overline{\sigma}_l )^\beta}_\delta
\ee

\be
  \delta_\alpha^\beta     (\sigma_\rho \overline{\sigma}_r)^i_{\ \ j}    (\sigma_\mu)^{\alpha \dot{\beta}}  (\overline{\sigma}_l)_{\dot{\beta} \gamma} \varepsilon^{\gamma j} \varepsilon_{i \delta} = {(\sigma_\mu \overline{\sigma}_l \sigma_r \overline{\sigma}_\rho)^\beta}_\delta
\ee
Combining all the equations above, we obtain

\bea\la{combine1}
\tilde{M}_\mu^{(1)} =- \int d^4 x \ \  \left( \frac{2 \rho^2}{ \lambda}   \sigma_\mu \overline{\sigma}_l   e^{ i p \cdot x} \frac{x^l}{(x^2+ \rho^2)^2} +  \sigma_\mu \overline{\sigma}_\rho k^\rho \frac{ \rho^4}{\lambda}( e^{i p \cdot x} - e^{- i q \cdot x} )  \frac{1}{(x^2+ \rho^2)^2 ( k \cdot x)}\right)
\eea
The $d^4 x$ integration in~\eqref{combine1}  can be done with the help of the following formula ($p^2 \longrightarrow 0$)

\be\la{dipvanish}
\int d^4 x \ \  e^{ i p \cdot x} \frac{x^l}{(x^2+ \rho^2)^2} =  i 2 \pi^2 \frac{ p^l}{p^2}
\ee

\be\la{formula1}
\int d^4 x    \frac{  e^{ i p \cdot x}}{(x^2+ \rho^2)^2 ( k \cdot x)} =  -i \frac{\pi^2}{p \cdot k} \frac{\rho |p|}{\rho^2} K_1  (\rho |p|)  =  i \frac{2 \pi^2}{q^2} \frac{\rho |p|}{\rho^2} K_1   (\rho |p| ) 
\ee
where we used $- 2 p \cdot k = (k - p)^2 -k^2 - p^2 \approx q^2$.   In our paper~\cite{Qian:2011ya}, we explicitly showed that all terms proportional to $\overline{\sigma}_l p^l$ vanish.  
Thus

\be\la{dipfinal1}
\tilde{M}_\mu^{(1)} =-  i \frac{4 \pi^2 \rho^2}{\lambda} \sigma_\mu  \overline{\sigma}_l   \frac{k^l}{q^2} [f(\rho |p|) - f(\rho |q|)]  
\ee
where $f(a) = a K_1 (a)$
As the incoming quark with momentum $p$ is on-shell and the mass of the quark is small ($p^2 \longrightarrow 0$), we have

\be
f(\rho |p|) = \rho |p| K_1 (\rho |p|) \longrightarrow \rho |p| \frac{1}{\rho |p|} = 1
\ee
Since $q^2 <0$ in SIDIS, we define $Q^2 = - q^2 > 0$. \eqref{dipfinal1} simplifies to

\be\la{dipfinal2}
\tilde{M}_\mu^{(1)} = i \frac{4 \pi^2 \rho^2}{\lambda} \sigma_\mu  \overline{\sigma}_l    \frac{k^l}{Q^2} \left[1 - f(\rho Q)\right]  
\ee

Here we note that~\eqref{dipfinal2} is derived from~\eqref{beforesimplify} which pictorially reads

\begin{itemize}
\item Left-handed quark ($\vec{p}$) $\xrightarrow{\rm Instanton}$   Right-handed quark (zero mode)   $\xrightarrow{{\rm Photon} \, \vec{q}}$ Right-handed quark ($\vec{k}$)
\end{itemize}
where $\vec{q} = \vec{k} -\vec{p}$.  On the other hand, if we consider
\begin{itemize}
\item Right-handed quark ($\vec{k}$) $\xrightarrow{{\rm Photon} \, - \vec{q}}$  Right-handed quark (zero mode) $\xrightarrow{\rm Anti-instanton}$ Left-handed quark ($\vec{p}$)
\end{itemize}
instead of~\eqref{dipfinal2}, we would obtain  
\be 
\tilde{M}_\mu^{(1)} = - i \frac{4 \pi^2 \rho^2}{\lambda^\dagger} \sigma_\mu  \overline{\sigma}_l    \frac{k^l}{Q^2} \left[1 - f(\rho Q)\right]  
\ee
where we have taken the conjugate of~\eqref{dipfinal2} and replaced $p \leftrightarrow k$. 
In~\cite{Qian:2011ya} we used the replacement $k \leftrightarrow - p$. We have checked that our 
final results are left unchanged by this correction. Thus

\bea\la{dipfinal3}
M_\mu^{(1)} =   4 \pi^2 \rho^2  \left(i \frac{\bold{P_+}}{\lambda} \sigma_\mu  \overline{\sigma}_l k^l - i\frac{\bold{P_-}}{\lambda^\dagger} \sigma_l  \overline{\sigma}_\mu p^l \right) \frac{[1 - f(\rho Q)]}{Q^2}   
\eea
where $\bold{P}_\pm=1/0$ denote one or no instanton/anti-instanton.  Similarly, for the processes depicted pictorially as

\begin{itemize}
\item Right-handed quark ($\vec{p}$) $\xrightarrow{\rm Anti-Instanton}$   Left-handed quark (zero mode)   $\xrightarrow{{\rm Photon} \, \vec{q}}$ Left-handed quark ($\vec{k}$)
 
\item Left-handed quark ($\vec{k}$) $\xrightarrow{{\rm Photon} \, - \vec{q}}$  Left-handed quark (zero mode) $\xrightarrow{\rm Instanton}$ Right-handed quark ($\vec{p}$)
\end{itemize}
Thus the result combining both the instanton and anti-instanton contributions

\bea\la{dipfinal4}
M_\mu^{(1)} =   4 \pi^2 \rho^2  \left(i \frac{\bold{P_+}}{\lambda} \gamma_\mu \slashed{k} - i\frac{\bold{P_-}}{\lambda^\dagger} \slashed{p} \gamma_\mu \right) \frac{[1 - f(\rho Q)]}{Q^2}   
\eea

Now, we need to average \eqref{dipfinal4} using the instanton liquid model. The standard averaging in the vacuum is

\be
\left<\frac{1}{\lambda} \right> = \frac{1}{2 N} \int \frac{d \lambda}{\lambda} \,  n (\lambda) = - i \pi \frac{n(0)}{N}
\ee
where by Banks-Casher relation is used $\pi n(0)/ N = -\left< q^\dagger q \right>$. 
However, we note that $\bold{P}_\pm$ in \eqref{dipfinal4} means that we fix an instanton or anti-instanton
pertaining to the polarized hadron prior to the averaging. This means that the pertinent eigenvalue distribution 
instead is

\be
n(\pm, \lambda) = n(\lambda) -\delta(\lambda \mp \lambda_* /N)
\ee
with $\lambda_*$ a typical eigenvalue in the zero-mode-zone.  Technically $n(\pm, \lambda)$ amounts to fixing an instanton or anti-instanton, and averaging over the remainder of the instanton-antiinstanton liquid by removing 1-row and 1-column in the $N \times N$ overlap matrix of zero-modes $\bold{T_{IJ}}$ for the fixed instanton or anti-instanton while averaging with $\textrm{det} \bold{T}$
in the instanton liquid model.  Explictly, this amounts to 

\be
\frac{\bold{P_\pm}}{\lambda} \xrightarrow{\rm fix \,\, an \,\, instanton/anti-instanton} \left< \frac{\bold{P_\pm}}{\lambda} \right> = \frac{1}{N} \int \frac{d \lambda}{\lambda} \,  n (\pm, \lambda) =  \mp \frac{1}{\lambda_*}  - i \pi \frac{n(0)}{N}
\ee 
Thus

\bea\la{dipfinal5}
\left< M_\mu^{(1)} \right> =   4 \pi^2 \rho^2   \frac{\pi n(0)}{N} \left(   \gamma_\mu \slashed{k} + \slashed{p} \gamma_\mu \right) \frac{[1 - f(\rho Q)]}{Q^2}    - i  \frac{4 \pi^2 \rho^2 }{\lambda_*} \left(   \gamma_\mu \slashed{k}+ \slashed{p} \gamma_\mu \right) \frac{[1 - f(\rho Q)]}{Q^2}   
\eea
The real part can be re-written as

\bea
4 \pi^2 \rho^2   \frac{\pi n(0)}{N} \left(   \gamma_\mu \slashed{k} + \slashed{p} \gamma_\mu \right) \frac{[1 - f(\rho Q)]}{Q^2}  &=&  4 \pi^2 \rho^2   \frac{\pi n(0)}{N} \frac{[1 - f(\rho Q)]}{Q^2} q^\nu \sigma_{\mu\nu} +  4 \pi^2 \rho^2   \frac{\pi n(0)}{N} \frac{[1 - f(\rho Q)]}{Q^2}   \left(   \slashed{k} \gamma_\mu -  \gamma_\mu \slashed{p}  \right) \nonumber\\
& \longrightarrow &4 \pi^2 \rho^2   \frac{\pi n(0)}{N} \frac{[1 - f(\rho Q)]}{Q^2} q^\nu \sigma_{\mu\nu}
\eea
where $\sigma_{\mu\nu} = [\gamma_\mu, \gamma_\nu]$. The parts proportional to $ \slashed{k} \gamma_\mu$ and $\gamma_\mu \slashed{p}$ vanish as discussed in~\cite{Kochelev:2003cp}. 
The imaginary part is

\be
 - i  \frac{4 \pi^2 \rho^2 }{\lambda_*} \left(   \gamma_\mu \slashed{k}+ \slashed{p} \gamma_\mu \right) \frac{[1 - f(\rho Q)]}{Q^2}   
\ee
and contributes to SSA in SIDIS as noted in~\cite{Ostrovsky:2004pd,Qian:2011ya}.

\subsection{\label{subsec:type2} Gluon vertex}

\begin{figure}
\includegraphics[height=40mm]{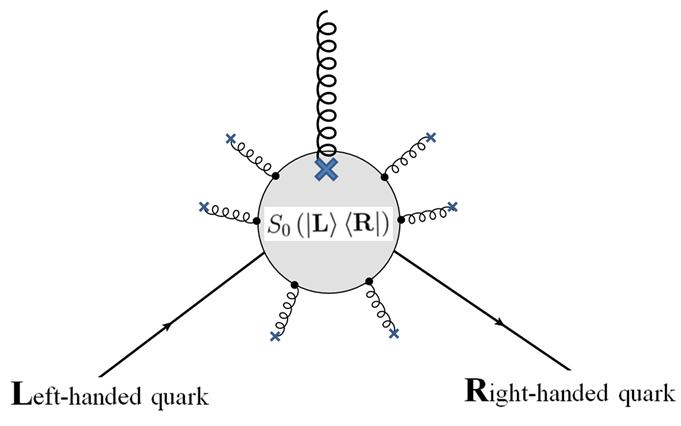}
\caption{ Effective Quark-Gluon vertex in the instanton vacuum.}
\label{backquantumvertex}
\end{figure}

The QCD vacuum is a random ensemble of  instantons and anti-instantons interacting via the exchange of
perturbative gluons and quasi-zero modes of light quarks and anti-quarks. In the dilute instanton approximation,
a typical effective vertex with quarks and gluons attached to an instanton is shown in Fig.~\ref{backquantumvertex}.
The corresponding effective vertex is given by~\cite{ 'tHooft:1976fv, Vainshtein:1981wh, Kochelev:1996pv},

\bea\la{antvertexeq1}
\mathcal{L} = \int \prod_q \left[ m_q \rho - 2 \pi^2 \rho^3 \bar{q}_R \left( 1 + \frac{i}{4} \tau^a \bar{\eta}_{\mu\nu}^a \sigma_{\mu\nu} \right) q_L \right] \exp\left( - \frac{2 \pi^2}{g_s} \rho^2 \bar{\eta}^b_{\gamma \delta} G^b_{\gamma \delta} F_g(\rho Q) \right) d_0 (\rho) \frac{d \rho}{\rho^5} d \bar{\sigma} + \left( L \leftrightarrow R \right)
\eea
where $d \bar{\sigma}$ is the integration over the instanton orientation in color space and  $\sigma_{\mu\nu} = [\gamma_\mu ,\gamma_\nu]/2$.  The incoming and outgoing quarks have small momenta $p$ ($\rho p\ll  1$) and $Q$ is the momentum
transferred by the inserted gluon with a form-factor

\be
F_g (x) \equiv \frac{4}{x^2} - 2 K_2 (x) \xrightarrow{x \rightarrow 0} 1
\ee
By expanding (\ref{antvertexeq1}) to leading order in the inserted gluon field 
of $G^b_{\gamma \delta}$ and integrating  over the color indices, we obtain

\be
    \frac{i}{g_s}  F_g(\rho Q)   \int   \pi^4 \rho^4 \frac{ \bar{q}_R   t^a   \sigma_{\mu\nu}   q_L }{m_q^*}   G^a_{\mu\nu} \times  \left( \prod_q \left( \rho m_q^* \right) d_0 (\rho) \frac{d \rho}{\rho^5} \right)  =   \frac{i}{g_s}    F_g(\rho Q)   \int   d \rho~\pi^4 \rho^4   n(\rho) \frac{ \bar{q}_R   t^a   \sigma_{\mu\nu}   q_L }{m_q^*}   G^a_{\mu\nu}   
\ee
where $n(\rho)$ is the effective instanton density and $m^*_q$ is the effective quark mass. 
In the dilute instanton approximation~\cite{Shuryak:1981fza}

\be
n(\rho) = n_I \delta (\rho - \rho_c)
\ee
where $\rho_c$ is the average size of the instanton.  Hence the induced instanton effective quark-gluon vertex

\be\la{antvertexeq2}
 \frac{i}{g_s}   F_g(\rho Q)   \pi^4   (n_I \rho^4_c)   \frac{ \bar{q}_R   t^a   \sigma_{\mu\nu}   q_L }{m_q^*}   G^a_{\mu\nu}   
\ee
as illustrated in Fig.~\ref{backquantumvertex}.  In momentum space, the effective vertex is $M_\mu^a$ and reads
\be  
\label{VERTEX}
M_\mu^a = t^a  \left[  \gamma_\mu-   \bold{P}_+ \gamma_+    \sigma_{\mu\nu} q^\nu    \Psi     -   { \bold{P}}_- \gamma_-     \sigma_{\mu\nu} q^\nu  \Psi   \right]
\ee
with $\gamma_\pm =(1\pm \gamma_5)/2$ and
\be
\Psi =   \frac{   F_g(\rho_c\, Q)   \pi^4   (n_I \rho^4_c)   }{m_q^* g_s^2}
\ee

The averaging of \eqref{VERTEX} in the instanton liquid gives

\be\la{effectivevertex}
\left<M_\mu^a \right> =  t^a  \left[  \gamma_\mu-       \sigma_{\mu\nu} q^\nu    \Psi   \right]
\ee
where we used

\be
\left< \bold{P}_+  \right> = \left< \bold{P}_-  \right> = 1
\ee
after the analytical continuation to Minkowski Space. (\ref{antvertexeq2}) yields an anomalously large  Quark Chromomagnetic Moment~\cite{Kochelev:1996pv}

\be
\mu_a = - \frac{2 n_I \pi^4 \rho_c^4}{g_s^2}
\ee

\bibliography{instantonref}

\end{document}